\documentclass[twocolumn]{aa} 
\usepackage{longtable}
\usepackage{graphicx}
\usepackage{txfonts}
\usepackage{csvsimple}
\usepackage{multirow}
\usepackage{rotating}
\usepackage{amsmath}
\usepackage{longtable}
\usepackage{booktabs}
\usepackage{ltablex}
\usepackage{epsfig}
\usepackage{threeparttable}
\usepackage{url}
\usepackage{color}
\usepackage{soul}
\usepackage[switch]{linenoaa}
\usepackage[dvipsnames]{xcolor}
\usepackage[colorlinks=true, linkcolor=MidnightBlue, citecolor=MidnightBlue, urlcolor=MidnightBlue]{hyperref}

\begin{document} 

   \title{In-flight calibration of the Wide-field X-ray Telescope on board the {\em Einstein Probe}}

   \author{Huaqing Cheng\inst{1}
          \and
          Hai-Wu Pan\inst{1}\fnmsep\thanks{Corresponding author: panhaiwu@bao.ac.cn}
          \and
          Yuan Liu\inst{1}\fnmsep\thanks{Corresponding author: liuyuan@bao.ac.cn}
          \and
          Jingwei Hu\inst{1}
          \and
          Haonan Yang\inst{1}
          \and
          Donghua Zhao\inst{1}
          \and
          Zhixing Ling\inst{1,2}
          \and
          Yifan Chen\inst{3}
          \and
          Xiaojin Sun\inst{3}
          \and
          Longhui Li\inst{4}
          \and
          Ge Jin\inst{4}
          \and
          Wenxin Wang\inst{1}
          \and
          Xue Yang\inst{1}
          \and
          He-Yang Liu\inst{1}
          \and
          Chen Zhang\inst{1,2}
          \and
          Shuang-Nan Zhang\inst{5,2}
          \and
          Weimin Yuan\inst{1,2}
          }

   \institute{Key Laboratory of Space Astronomy and Technology, National Astronomical Observatories, Chinese Academy of Sciences, Datun Road 20A, Chaoyang District, Beijing 100101, China
         \and
             School of Astronomy and Space Science, University of Chinese Academy of Sciences, Beijing 100049, China
        \and
            Shanghai Institute of Technical Physics, Chinese Academy of Sciences, Yutian Road 500, Shanghai 200083, China
        \and
            North Night Vision Technology Co., LTD, Nanjing, 211106, China
        \and
            State Key Laboratory of Particle Astrophysics, Institute of High Energy Physics, Chinese Academy of Sciences, Beijing 100049, China}

  \abstract
 {By utilizing novel lobster-eye optics, the Wide-field X-ray Telescope (WXT) onboard the \textit{Einstein Probe} (EP) satellite achieves an unprecedented combination of a large instantaneous field-of-view (FoV) and high sensitivity for monitoring the dynamic X-ray sky. In this paper, we present the in-orbit calibration results of the WXT during its first two and a half years of operations. By conducting observations of standard celestial sources---including the Crab Nebula, Scorpius X-1, and Cassiopeia A---we systematically characterized key instrumental properties. Our analysis demonstrates that the in-orbit performance of the WXT agrees with prelaunch ground calibrations well. The spatial resolution, denoted by the full width at half maximum (FWHM) of the focal spot, typically ranges from $3'$ to $6'$ across $\sim$90\% of the FoV, with a median of $\sim 4.3'$. The post-calibration source positioning accuracy achieves $1.3'$ (at the 90\% confidence level). The in-orbit effective area is consistent with model predictions and ground measurements, exhibiting an overall systematic uncertainty of $\lesssim 10\%$ (90\% C.L.) in the 0.5--4\,keV band. While the vast majority of the detectors remain highly stable, a noticeable long-term degradation at the low-energy end ($\sim30\%$--$40\%$, 0.4--0.6 keV) is observed in a few specific modules. Furthermore, spectral evaluations using Cas A confirm the stability of the energy scale and spectral resolution of the focal-plane Complementary Metal-Oxide Semiconductor (CMOS) detectors. All derived calibration products have been incorporated into the WXT calibration database (CALDB). These results comprehensively verify the instrumental capabilities of the WXT, providing a solid foundation for the reliable analysis of scientific observations.}

               
   \maketitle

\section{Introduction}
\label{sec:intro}

The dynamic X-ray sky is rich with transient events and variable sources, spanning a broad spectrum of timescales and brightness levels. To effectively monitor this cosmic domain requires instruments with extraordinarily large instantaneous fields of view (FoV) and high sensitivity. Building upon the legacy of previous generations of wide-field X-ray monitors, the \textit{Einstein Probe} \citep[\textit{EP};][]{2016SSRv..202..235Y,Yuan2018,Yuan2022, Yuan2025}, an interdisciplinary observatory dedicated to time-domain astronomy and high-energy astrophysics, was launched on January 9, 2024. The \textit{EP} mission is led by the Chinese Academy of Sciences (CAS) in collaboration with the European Space Agency (ESA), the Max Planck Institute for Extraterrestrial Physics (MPE), and the French Space Agency (CNES). Following a successful commissioning and instrumental calibration phase, \textit{EP} commenced nominal scientific operations in July 2024. Its first two and a half years of operations have yielded a wealth of scientific discoveries across various fields of high-energy astrophysics, demonstrating its impressive capabilities for X-ray sky surveys. These encompass the discoveries of $\sim200$ fast X-ray transients \citep[e.g., gamma-ray bursts, X-ray flashes, and supernova shock breakouts;][]{Yin2024, LiuY2025, 2024HSun, Jiang2025, 2025LiWX, 2025EPZTFSN, 2025YinYihan, 2026DaiGRB,2026YuanEPSN}, tidal disruption events \citep[TDEs, e.g.,][]{2025Jin, 2025ShuXinwen, 2025Li0702}, new candidate X-ray binaries \citep[e.g.,][]{2025HaberlXu,EP240904a,2026WangYLXRB, 2026HuangGuoli, 2026FrancescoEPBH}, outbursts from known accreting systems \citep[e.g.,][]{Marino2025, 2025YangBeXRB, 2025LiZhaosheng}, X-ray flares from active galactic nuclei \citep[AGNs, e.g.,][]{LiuMJ_AGN, 2026LuobinEPAGN}, and transients of an undetermined nature \citep[e.g.,][]{ZhangWD2025}.

Onboard the \textit{EP} satellite are two primary scientific payloads: the Wide-field X-ray Telescope \citep[WXT;][]{2025ChengWXTcalib}, featuring novel lobster-eye micro-pore optics \citep[MPO, e.g.,][]{Angel1979,Fraser1992, Fraser1993,Willingale1998,Willingale2016}, and the Follow-up X-ray Telescope \citep[FXT;][]{2020SPIE11444E..5BC}, a standard Wolter-I type telescope. As the principal payload for discovering transients and monitoring source variability in the soft X-ray band (0.5--4\,keV), the \textit{EP}-WXT combines an unprecedentedly large instantaneous FoV of $>3600$ square degrees with high detection sensitivity, achieving a flux limit of $\sim (2-3)\times10^{-11}~{\rm erg~s^{-1}~cm^{-2}}$ in a typical $\sim 1000$\,s exposure. The instrument comprises twelve identical flight model (FM) modules (labeled FM 1--12), each covering a FoV of $18.6^\circ\times18.6^\circ$. For the focal plane detectors, large-format, back-illuminated scientific complementary metal-oxide semiconductor \citep[CMOS, e.g.,][]{2022WangWXa,2022WangWXb,Wu2022,Wu2023pasp1,Wu2023pasp2,Wu2023nima,ChenMX2024, LiuMJ2025} sensors are employed, leveraging their high readout speed and significantly lower cost compared to traditional charge-coupled devices (CCDs). A single CMOS sensor consists of $4096\times4096$ pixels, with a pixel scale of $8.25''$. Each module integrates $3\times3$ MPO plates and four CMOS sensors\footnote{The CMOS numbering follows the module sequence, i.e., CMOS~1--4 for FM~1, CMOS~5--8 for FM~2, and so on.}. To the best of our knowledge, the \textit{EP}-WXT represents the first implementation of mass-produced MPO optics and CMOS detectors in a dedicated X-ray astronomy mission.

For a thorough and in-depth characterization of the instrumental properties, tremendous efforts have been devoted to both the ground and in-flight calibrations of the WXT payload. Specifically, the ground calibration campaigns were meticulously designed, spanning over two years given the large number of individual modules and the demanding schedule of the development phase (Phase D). During this period, a series of tests and experiments were conducted to calibrate the physical properties at the levels of devices \citep[e.g.,][]{2022PASP..134k5002L, LI2024130821, Wu2023pasp2}, assemblies \citep[e.g.,][]{Suri_LEIA,Rukdee_EPWXT, Chenyifan2026}, and complete modules \citep{2025ChengWXTcalib}. The ground calibration database (CALDB), constructed from the results of these experiments, was then fully incorporated into the scientific data analysis pipeline after the launch of \textit{EP}. Despite these extensive prelaunch endeavors, the implementation of certain calibration terms was hindered by ground-based experimental limitations. For instance, the calibration of source positioning could not be performed prior to launch, making it a primary objective during the mission's commissioning phase. Establishing this initial in-orbit baseline---particularly achieving a positional accuracy of $\le 2'$ alongside comprehensive characterizations of the point-spread function (PSF), effective area, and energy response---serves as the fundamental prerequisite for precise and reliable scientific analysis. Furthermore, beyond this initial commissioning, certain instrumental properties (e.g., the low-energy effective area and detector energy response) are subject to long-term evolution. These variations are induced by a variety of space-environment factors, such as contamination accumulation \citep[e.g.,][]{Marshall2004,Plucinsky2022}, variations in the space radiation environment \citep[e.g.,][]{Grant2005, O_Dell2007}, and micrometeoroid impacts \citep[e.g.,][]{Abbey2006, Carpenter2008}. Consequently, these performance metrics require continuous tracking and validation through periodic in-orbit calibration procedures.

It should be emphasized that the in-orbit calibration of the \textit{EP}-WXT is fundamentally built upon a previous calibration campaign implemented on its pathfinder mission, the Lobster Eye Imager for Astronomy \citep[\textit{LEIA};][]{2023LingZXRAA}. Carrying a qualification model (QM) of one of the twelve identical modules of the \textit{EP}-WXT instrument, \textit{LEIA} was launched on July 27, 2022. During its two and a half years of operations, \textit{LEIA} successfully verified the in-orbit performance of the MPO and CMOS technologies \citep[e.g.,][]{2022ZhangChenApJL, 2025SunNSR, 2025MaoSF, 2025YangBeXRB}, and rehearsed the in-flight calibration protocols for such wide-field lobster-eye X-ray telescopes \citep{2025ChengLEIA}. We note, however, that the calibration of the \textit{EP}-WXT payload is not a simple duplication of \textit{LEIA}'s operations; rather, it introduces a significant leap in both scale and complexity. First, as a dedicated technology demonstration mission, \textit{LEIA} enjoyed substantial flexibility in scheduling extensive calibration observations. In contrast, \textit{EP} must balance the comprehensive calibration of its twelve modules against a highly demanding scientific mandate. The mission's primary objectives---such as the continuous all-sky monitoring for X-ray transients and the critical search for electromagnetic counterparts of gravitational wave and neutrino event signals---consume the vast majority of operational time, thereby severely constraining the temporal windows available for routine calibration. Second, scaling from a single QM to twelve flight modules (FMs) necessitates the rigorous verification of consistent optical performance (e.g., PSF, source positioning, effective area) across an unprecedented $>3600$ square degree FoV. Furthermore, subjecting a massive array of 48 ($4\times12$) commercial-off-the-shelf-like scientific CMOS sensors to the harsh low-Earth orbit environment over a multi-year period is without historical precedent in X-ray astronomy. Monitoring their long-term performance evolution and radiation degradation is not only critical for \textit{EP}, but also provides an invaluable benchmark for future wide-field X-ray missions.

During the first two and a half years of on-orbit operations, we have carried out three rounds of in-orbit calibrations. In this paper, we present the results of these calibrations in detail, focusing on several key instrumental properties: the PSF, source positional accuracy, effective area, and energy response (energy-channel relation and spectral resolution). The structure of this paper is organized as follows. Section \ref{sec:calib_procedure} reviews the main procedures of the calibration campaigns implemented so far. We present the calibration results of the PSF in Sect.~\ref{sec:psf}, source positioning in Sect.~\ref{sec:source_positioning}, the effective area in Sect.~\ref{sec:effarea}, and the energy response in Sect.~\ref{sec:cmos_energy}. The summary and conclusions are given in Sect.~\ref{sec:summary}.

\section{The calibration campaigns}
\label{sec:calib_procedure}

\begin{table*}[hbtp]
\centering
\caption{Specifications of the \textit{EP}-WXT instrument.}
\label{tab:epwxt_specification}
\begin{tabular}{l c c}
\toprule
Parameter & Goal & Ground measurements \\
\midrule
Angular resolution (FWHM, $\sim 1$\,keV) & $\le 5'$ & $\sim 3'$--$5'$ \\ 
Source positioning & $\le 2'$ (J2000, 90\% C.L.) & --- \\
Effective area (focal spot, 1\,keV) & $\ge 2.7$\,cm$^2$ & $\sim 3$\,cm$^2$ \\ 
Energy band & 0.5--4\,keV & $\sim 0.4$--$6$\,keV \\ 
Energy resolution (1.25\,keV) & $< 170$\,eV & $\sim 120$--$140$\,eV \\
Field of view (FoV) & $> 3600$ square degrees & $\approx 3850$ square degrees \\
\bottomrule
\end{tabular}
\end{table*}

\subsection{Main goals}
\label{sec:wxt_goal}

Table~\ref{tab:epwxt_specification} summarizes the specifications of the \textit{EP}-WXT payload, encompassing both the design requirements and the results derived from ground calibrations. As the primary scientific payload of the \textit{EP} satellite, the WXT is designed to perform a high-cadence survey of the soft X-ray sky by combining a large instantaneous field of view (FoV) with high sensitivity. Its main scientific objectives are to discover new X-ray transients and to monitor the long-term variability of known X-ray sources \citep{Yuan2025}.

Accurate source localization at the level of $\sim 1'$--$2'$, which is primarily governed by the imaging performance, is essential for enabling efficient follow-up observations both with the onboard \textit{EP}-FXT payload and with external multi-wavelength facilities. Reliable measurements of source fluxes and spectral properties, which depend crucially on the calibration of the effective area and energy response, are vital for characterizing the radiation mechanisms and physical origins of the observed sources. 

This paper focuses on the in-orbit calibration of these key spatial and spectral properties, which are most relevant to general scientific investigations and constitute the primary objectives of the current calibration campaigns. Other specific aspects fall beyond the scope of this study and will be detailed in separate publications. For instance, the timing accuracy of the CMOS sensors is currently undergoing rigorous cross-calibration and will be presented in a dedicated future paper (Cheng et al. in prep.). Similarly, the secular evolution of the detectors---such as instrumental background properties and performance variations (e.g., changes in the bias map or newly emerged bad pixels)---is also planned to be published elsewhere (Ling et al., in prep.).

Another critical consideration during in-orbit operations is the impact of micrometeoroids and space debris, which can induce detector damage (e.g., the emergence of bright columns or bad-pixel clusters) and performance degradation (e.g., a reduction in effective area), representing a significant risk for X-ray missions \citep[e.g.,][]{Drolshagen2019,YangXue2022}. Such impacts have been well documented in previous X-ray missions, such as those of the EPIC-pn camera onboard \textit{XMM-Newton} and the X-ray Telescope (XRT) onboard the \textit{Neil Gehrels Swift Observatory} \citep[e.g.,][]{Struder2001,Abbey2006,Carpenter2008}. The \textit{EP} satellite has experienced $\sim 20$ suspected impacts since its launch. For most events, the effects on the CMOS sensors were minor, with typically only $\sim 100$ bright pixels (note: the full-scale sensor comprises $4096\times4096$ pixels) emerging post-impact. These affected pixels were successfully identified, masked, and incorporated into the updated bad-pixel map within the CALDB. However, two detectors (CMOS 34 and CMOS 41) suffered moderate damage shortly after launch (on April 30, 2024, and May 7, 2024, respectively). These detectors subsequently exhibited notable variations in both the effective area and the energy-channel (E-C) relations across the detector plane, necessitating specialized calibration procedures and modeling. A comprehensive investigation into the micrometeoroid impacts on the \textit{EP} mission will be presented elsewhere (Yang et al., in prep.). For the purposes of this study, results from these two specific detectors are excluded from the general discussion regarding effective area and energy response calibrations.

\subsection{Calibration process and targets}
\label{sec:calib_process}

\begin{table*}[hbtp]
\caption{Observational log of the three in-orbit calibration campaigns for the \textit{EP}-WXT payload.} 
\label{table:calibplan}
\centering
\begin{tabular}{c l c l l c}
\toprule
Round & Target & Observation date & Calibrated modules & Calibration term & Test points / detector \\
\midrule
\multirow{3}{*}{1} 
 & Cas A & 2024.1 & 4, 10 & Gain, Energy resolution & 1 \\
 & Crab & 2024.1--3 & 1--2, 4--8, 10--12 & PSF, Positioning, Effective area & 36/37/41$^\star$ \\
 & Sco X-1 & 2024.5 & 3, 9 & PSF, Positioning & 36 \\
\midrule
2 & Crab & 2025.1--2 & 12 modules & Effective area & 1 \\
\midrule
\multirow{2}{*}{3} 
 & Cas A & 2025.8--12 & (-)$^\dagger$ & Gain, Energy resolution & - \\
 & Crab & 2025.10--2026.3 & 12 modules & Effective area & 1 \\
\bottomrule
\end{tabular}
\vspace{1ex}
\begin{minipage}{0.95\linewidth}
\small
\textbf{Notes:} \\
$^\star$ A standard sampling grid consisting of $6\times6$ directions was adopted for 38 detectors (CMOS 1--4, 9--12, 13, 15--28, 33--36, 38--48). Additional on-axis observations were employed for CMOS 14 and 37. Additional on-axis observations and four complementary points were introduced for the eight detectors onboard FMs 2 and 8 (CMOS 5--8, 29--32). \\
$^\dagger$ For regular operational modes (a solar avoidance angle of 94.5$^\circ$), the visibility window for Cas A is highly constrained. For a few detectors unable to observe this standard target, spectral analysis of the Crab nebula was alternatively employed to indirectly verify their energy response properties.
\end{minipage}
\end{table*}

The \textit{EP} satellite has been operating in orbit for about two and a half years, during which we have initiated three dedicated rounds of in-orbit calibration campaigns. Specifically, the first round of calibration was carried out during the commissioning phase, spanning approximately five months from January to May 2024. Its primary objective was to provide a comprehensive characterization of the in-orbit performance and instrumental baseline following the launch. Three well-established celestial calibration sources---the Cassiopeia A (Cas A) supernova remnant, the Crab nebula, and the bright, stable X-ray binary Scorpius X-1---were selected. These targets were scheduled for observation with specific detector modules, strategically accommodating the limited visibility windows inherent to the commissioning phase. 

Specifically, Cas A was observed in late January 2024 using two modules (4 and 10) for energy-response calibration. We subsequently observed the Crab nebula using ten modules (all except 3 and 9) to characterize the PSF and effective area, as well as to perform source positioning calibrations. Because the Crab nebula was unobservable when scheduling the remaining two modules (3 and 9), we alternatively targeted Sco X-1 in May 2024 to calibrate their PSF performance and source localization.

During the commissioning phase, the Cas A observations were performed solely targeting the detector center\footnote{The spatial variation of the energy response parameters (i.e., E-C relation and spectral resolution) has been found to be marginal across a specific detector plane. In general, for a given CMOS sensor, the gain coefficient varies within $1\%$ and the energy resolution within $\sim 5\%$ \citep[e.g.,][]{Cheng2024, 2025ChengWXTcalib}.}. For the other two calibration targets (the Crab and Sco X-1), a standard sampling grid of $6\times6$ evenly distributed incident angles (i.e., 36 points) was adopted for 38 detectors. For two specific detectors (CMOS 14 and 37) whose FoVs overlap with the FXT payload, an additional observation was performed toward the detector center, yielding 37 points. For the remaining eight detectors (CMOS 5--8 and 29--32), both the central observation and four supplementary pointing directions were introduced to further constrain the imaging properties, resulting in a total of 41 points per detector. This wide range of incident angles sampled across the expansive FoV enabled the robust derivation of the in-orbit transformation matrix between the celestial and detector coordinate systems, while simultaneously verifying the spatial uniformity of the optical performance (i.e., PSF and effective area) as previously indicated by ground calibration experiments \citep{2025ChengWXTcalib}.

A second in-orbit calibration campaign was conducted from January to February 2025, primarily utilizing observations of the Crab to investigate the year-scale stability of the effective area. The third calibration campaign commenced with observations of Cas A in August 2025 and concluded with those of the Crab in early 2026. The detailed schedule of these in-flight calibration observations is summarized in Table~\ref{table:calibplan}.

\subsection{Data reduction}
\label{sec:data_reduction}

The \textit{EP}-WXT produces two types of scientific data. The event data record the detected X-ray events and are primarily used for scientific research, whereas raw detector images can also be recorded to periodically examine the status of the CMOS sensors and the performance of the optical blocking filters. The data handled in this work are exclusively event data taken in photon counting (PC) mode. 

For each observation, the X-ray events were processed and calibrated using the official data analysis software (Liu et al., in prep.) and the calibration database \citep[CALDB;][]{2025ChengWXTcalib} designed for the \textit{EP} mission. The energy of each event was corrected using the bias and gain maps stored in the ground CALDB, and bad or flaring pixels were systematically flagged. Subsequently, the position of each valid event was projected into celestial coordinates (J2000). 
Single-, double-, triple-, and quadruple-pixel events without anomalous flags were selected to construct the cleaned event files. From these files, the light curves and spectra of both the source and the background were extracted. Source photons were extracted from a circular region centered on the target with a radius of $9.2'$. For the extraction of background photons, a source-centered annular region was employed, with inner and outer radii of $18.2'$ and $36.4'$, respectively. The ancillary response files (ARFs) and redistribution matrix files (RMFs) were also generated for the spectral analysis. Specifically for the analysis of the Crab nebula, the source spectra were re-binned to ensure a minimum of 25 photons per energy bin using the \texttt{grppha} tool (version 3.0.1). 

\begin{figure*}[!htbp]
    \centering
    \includegraphics[width=0.8\textwidth]{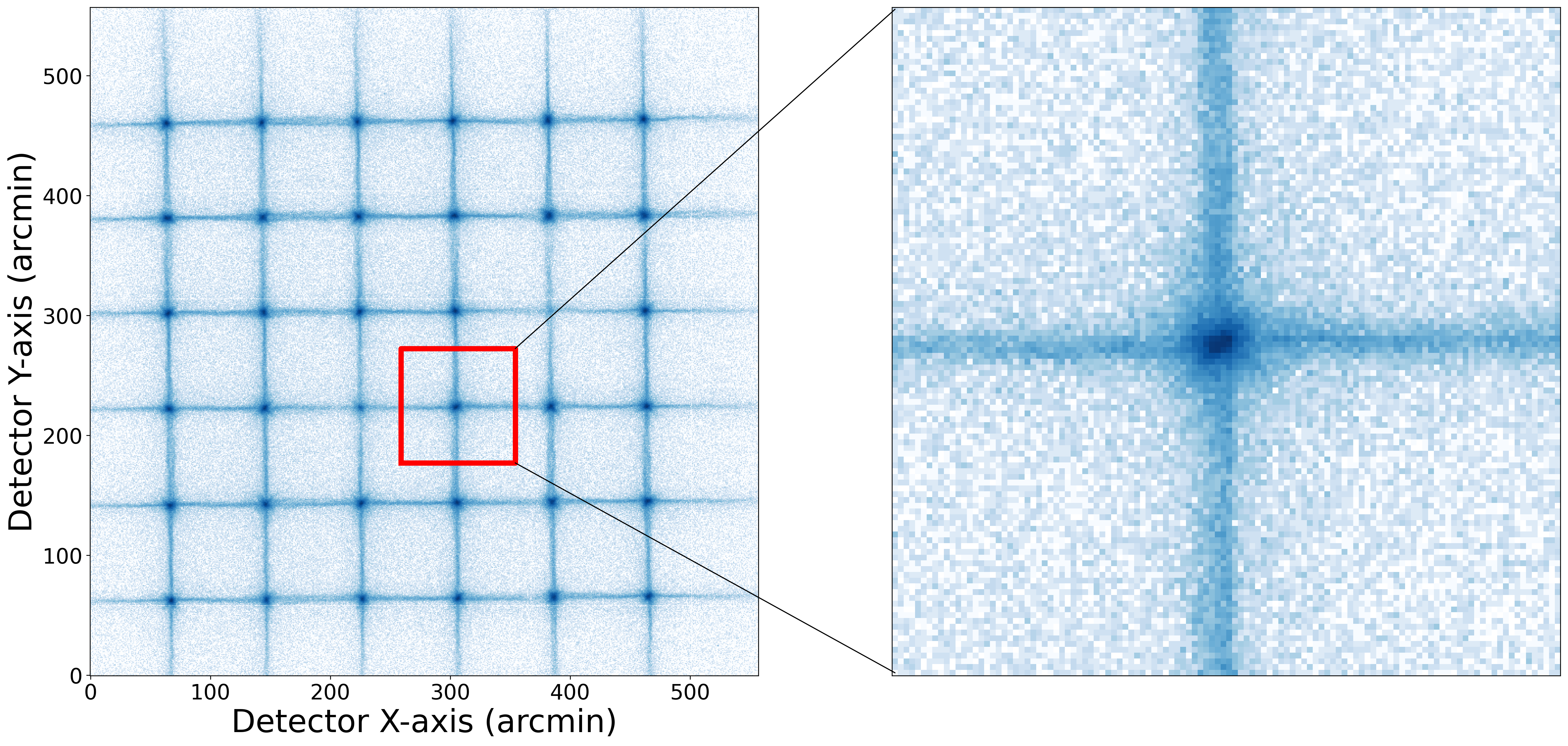}
    \caption{({\it Left} panel) Mosaic of observational images of the Crab nebula taken from $6\times6$ incident directions across the FoV of CMOS 1. ({\it Right} panel) Zoom-in image of the PSF sampled at the detector coordinates of [RAWX, RAWY] $\approx$ [2250, 1650]. The PSF is composed of a bright focal spot region and two orthogonal cruciform arms, as predicted by lobster-eye optics.}
    \label{fig:crabscan}
\end{figure*}

\begin{figure*}[!htbp]
    \centering
    \includegraphics[width=0.8\textwidth]{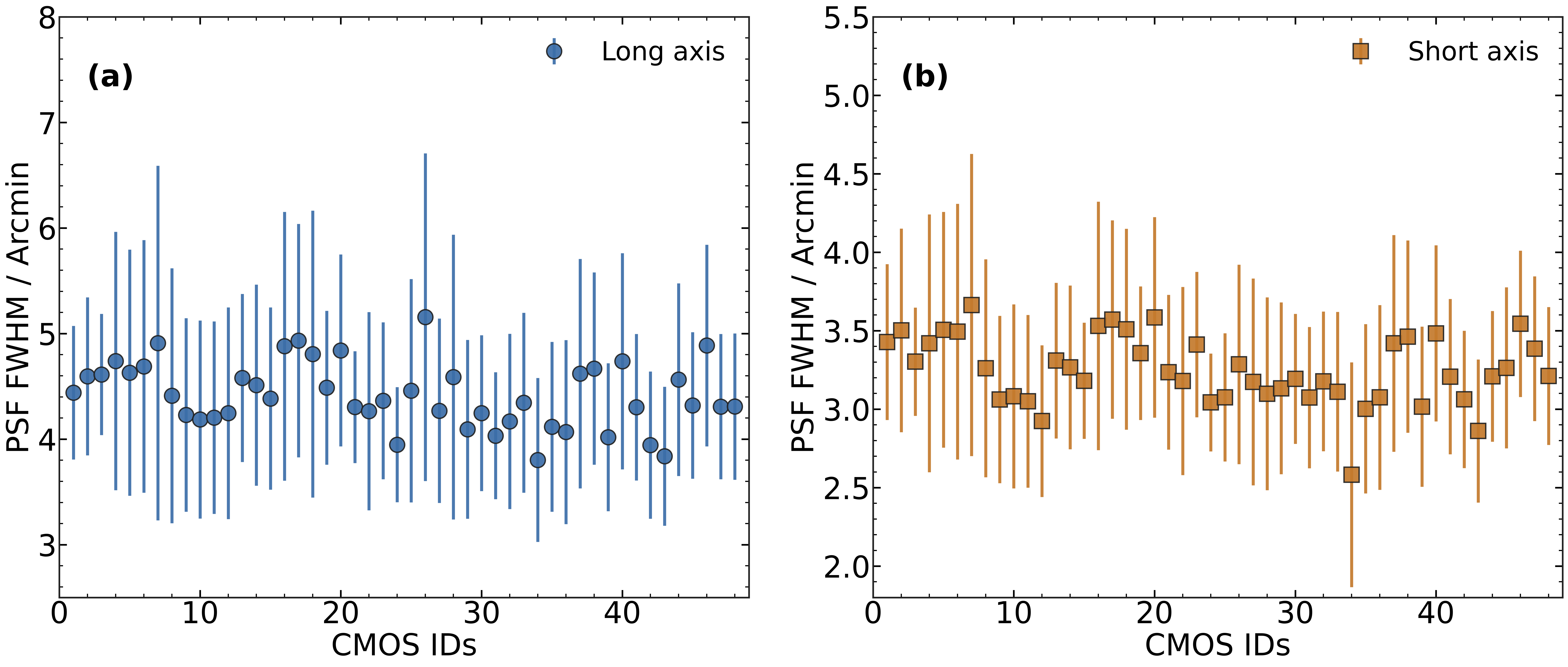}
    \caption{In-orbit PSF FWHM measured across different CMOS detectors. The \textit{left} panel shows the mean and standard deviation of the FWHM measured along the major axis. The \textit{right} panel shows the measurements along the minor axis.}
    \label{fig:psf_fwhm_each_cmos}
\end{figure*}

\begin{figure*}[!htbp]
    \centering
    \includegraphics[width=0.9\textwidth]{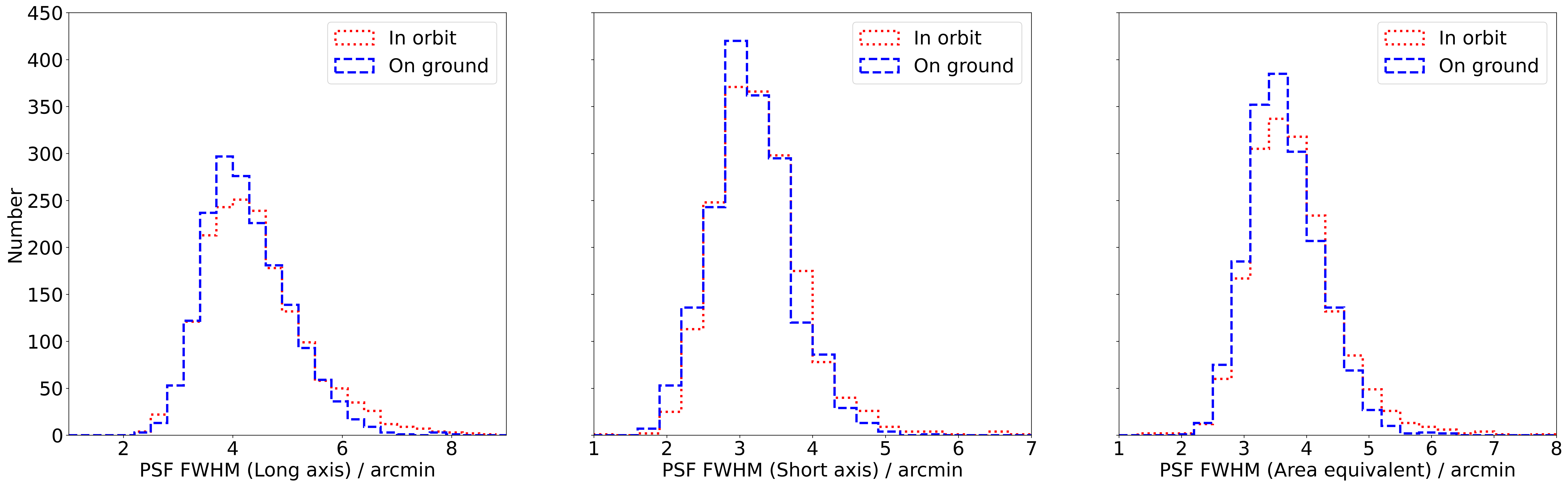}
    \caption{Distributions of the three characteristic measurements of the PSF FWHM obtained from the Crab and Sco X-1 observations during the first stage of calibration. From {\it left} to {\it right} panels: the lengths of the major axis, minor axis, and the area-equivalent radius (see text for detailed definitions). The ground measurements obtained at the XIB test facility (NAOC/CAS) are overplotted for comparison.}
    \label{fig:fwhm_variation_3types}
\end{figure*}

\section{Point-spread function}
\label{sec:psf}

The PSF calibration was performed through the image analysis of two bright, stable X-ray sources: the Crab (for the 40 detectors onboard FMs 1, 2, 4, 5, 6, 7, 8, 10, 11, and 12) and Sco X-1 (for the eight detectors onboard FMs 3 and 9). It should be noted that the Crab can be effectively treated as a point source for this analysis, as its emission in the WXT energy band is strongly dominated by the central pulsar \citep[e.g.,][]{2015Mason_NuSTAR_calibration, Nakajima_Hiroshi_Hitomi2018, 2025ChengLEIA}.
For the majority of the 48 CMOS detectors (38 modules), we adopted a standard sampling array consisting of $6\times6$ evenly distributed incident directions. For the two detectors whose FoVs overlap with the \textit{EP}-FXT payload (CMOS 14 and 37), an additional on-axis pointing was performed. For the remaining eight detectors aboard FMs 2 and 8 (i.e., CMOS 5--8 and 29--32), both the central observation and four supplementary sampling points were introduced, resulting in a total of 41 directions across their respective FoVs.
As a representative example, the mosaic image composed of the 36 pointed observations of the Crab for CMOS 1 is presented in the \textit{left} panel of Fig.~\ref{fig:crabscan}.
The cruciform morphology of the PSF exhibits excellent spatial uniformity across the FoV, which is in good agreement with the ground measurements obtained at the XIB test facility of NAOC (Zhang et al., in prep.), as well as those obtained at the 100-meter X-ray facility of IHEP \citep{2025ChengWXTcalib}.

To illustrate the detailed structural morphology of the PSF, the \textit{right} panel of Fig.~\ref{fig:crabscan} displays a zoom-in image taken at the detector position of [RAWX, RAWY] $\approx$ [2250, 1650]. 
As theoretically predicted for lobster-eye optics, the PSF is composed of a prominent central focal spot accompanied by two orthogonal cruciform arms. 
The former is produced by photons undergoing an even number of reflections (predominantly two) within the orthogonal micro-pore channel walls, while the latter originates from photons experiencing an odd number of reflections (predominantly one) \citep[e.g.,][]{Angel1979, Fraser1992}.
It is worth noting that the PSF structures observed across other CMOS sensors are highly consistent.
In particular, no significant morphological misalignment is observed across the entire vast FoV of the \textit{EP}-WXT instrument. We thus conclude that all 432 MPO plates onboard the \textit{EP}-WXT were mounted precisely with respect to the designed spherical focal surface.

Following the standard procedure for PSF characterization \citep[see Sect.~3.1 of][]{Cheng2024}, an elliptical function was employed to fit the half-maximum contour of the focal spot region, deriving the full width at half maximum (FWHM) of the PSF. From this fit, three characteristic PSF metrics were extracted: the lengths of the major and minor axes of the ellipse, and the area-equivalent radius, defined as the square root of the product of the major and minor axes. 
For each of the 48 CMOS detectors, the mean value and standard deviation of the PSF FWHM obtained from different sampling directions were calculated and are summarized in Fig.~\ref{fig:psf_fwhm_each_cmos}.
The \textit{left} panel displays the FWHM measured along the major axis, and the \textit{right} panel shows the results along the minor axis. 
The imaging quality across the various detectors is found to be largely consistent, with a mean major-axis FWHM of $\sim 4'$--$5'$ and a mean minor-axis FWHM of $\sim 2.5'$--$3.5'$.
Furthermore, for the majority of the CMOS sensors, we observe a remarkable uniformity across different incident angles, as evidenced by a relatively small intrinsic dispersion of $\sigma({\rm FWHM})\lesssim 1'$.
This effectively validates the spatial uniformity of the PSF characteristics across the detector FoV, in good agreement with the theoretical expectations for lobster-eye optics.

In addition, to investigate any potential deterioration of the PSF after launch, these three in-orbit FWHM metrics were compared with those obtained from the ground calibration campaigns, as presented in Fig.~\ref{fig:fwhm_variation_3types}. Clearly, no discernible degradation of the spatial resolution is observed post-launch. The in-flight spatial resolution of the WXT typically ranges from $3'$ to $6'$ for the vast majority ($\sim 90\%$) of the FoV, with a median value of $\sim 4.3'$ \citep[adopting the major-axis length as the primary indicator of spatial resolution following our previous convention;][]{Cheng2024, 2025ChengWXTcalib}, which fully meets the mission design requirements. Furthermore, the second and third rounds of PSF calibrations revealed no significant temporal variations in the PSF characteristics over the initial two and a half years of operations.

\begin{figure*}[!htbp]
    \centering
    \includegraphics[width=0.9\textwidth]{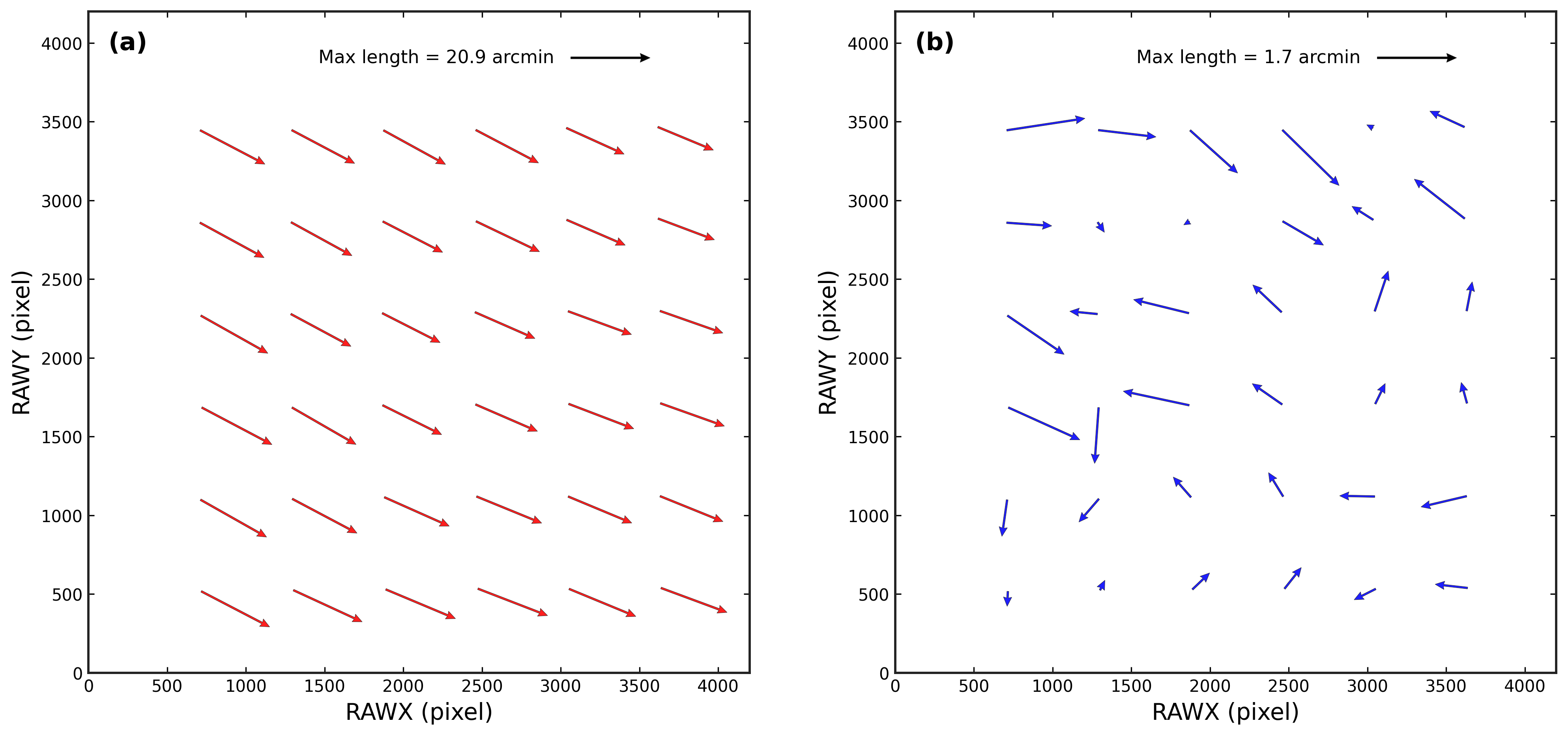}
    \caption{Exemplary quiver map representing the positional deviations between the expected and measured PSF focal spot centroids, using Sco X-1 data observed with CMOS 9. Panel (a) illustrates the offsets prior to the in-orbit localization calibration, and panel (b) shows the residual deviations after the calibration was applied. In each panel, a reference arrow representing the maximum offset vector is provided for scale.}
    \label{fig:quivermap}
\end{figure*}

\section{Source positioning calibration}
\label{sec:source_positioning}

The primary scientific goal of the \textit{EP}-WXT instrument is to discover new X-ray transients and to monitor the long-term X-ray activity of known celestial objects, leveraging its unprecedented combination of an expansive instantaneous FoV and high detection sensitivity \citep{Yuan2025}. 
Precise source localization is essential for enabling prompt and efficient follow-up observations with both the onboard \textit{EP}-FXT and external multi-wavelength facilities. Therefore, positional calibration constituted a major task during the commissioning phase. In practice, the source localization calibration involves refining the celestial-to-detector coordinate transformation matrices and modeling the residual nonlinear positional offsets. A detailed description of the calibration methodology can be found in Sect.~4 of \citet{2025ChengLEIA}. In this section, we briefly summarize the main calibration procedures and the corresponding results. 

For each detector, which covers an individual $9.3^\circ\times 9.3^\circ$ FoV, a series of pointed observations of the PSF calibration sources (the Crab for ten modules and Sco X-1 for the remaining two) was conducted. Through these observations, a largely uniformly distributed sample of PSFs at various incident angles was constructed. For each pointed observation, the theoretical position of the focal spot on the detector was derived by employing the satellite attitude data, the ground-measured positions of the curvature centers, and the initial celestial-to-detector transformation matrices. These theoretically expected positions were then compared with the actual measured PSF positions derived from the focal spot analysis, resulting in the construction of a quiver map of positional offsets.

Prior to the calibration efforts, large systematic positional deviations of up to $\sim 20'$ were observed (see Fig.~\ref{fig:quivermap}a). This was anticipated, as no end-to-end localization calibration could be implemented during the ground-testing stage due to experimental limitations. To mitigate these prominent offsets, an additional rotational matrix parameterized by three Euler angles was introduced. 
This additional matrix accounts for potential post-launch variations in the transformation matrices among different instrument components, as well as intrinsic pointing uncertainties in the star tracker system. With this correction applied, the positional deviations were significantly reduced, exhibiting a maximum offset value of $\sim 2'$ alongside a largely anisotropic residual pattern across the FoV (see Fig.~\ref{fig:quivermap}b).
These remaining nonlinear offsets are attributed primarily to intrinsic geometrical imperfections in the MPO plates introduced during the manufacturing process. 
They were thus modeled using the \texttt{scipy.interpolate.RBFInterpolator} algorithm via \texttt{SciPy} \citep{2020SciPy-NMeth} and stored as a $257\times257$ spatial correction matrix in the updated CALDB, alongside the refined celestial-to-detector transformation matrices.

\begin{figure}[!htbp]
    \centering
    \includegraphics[width=\columnwidth]{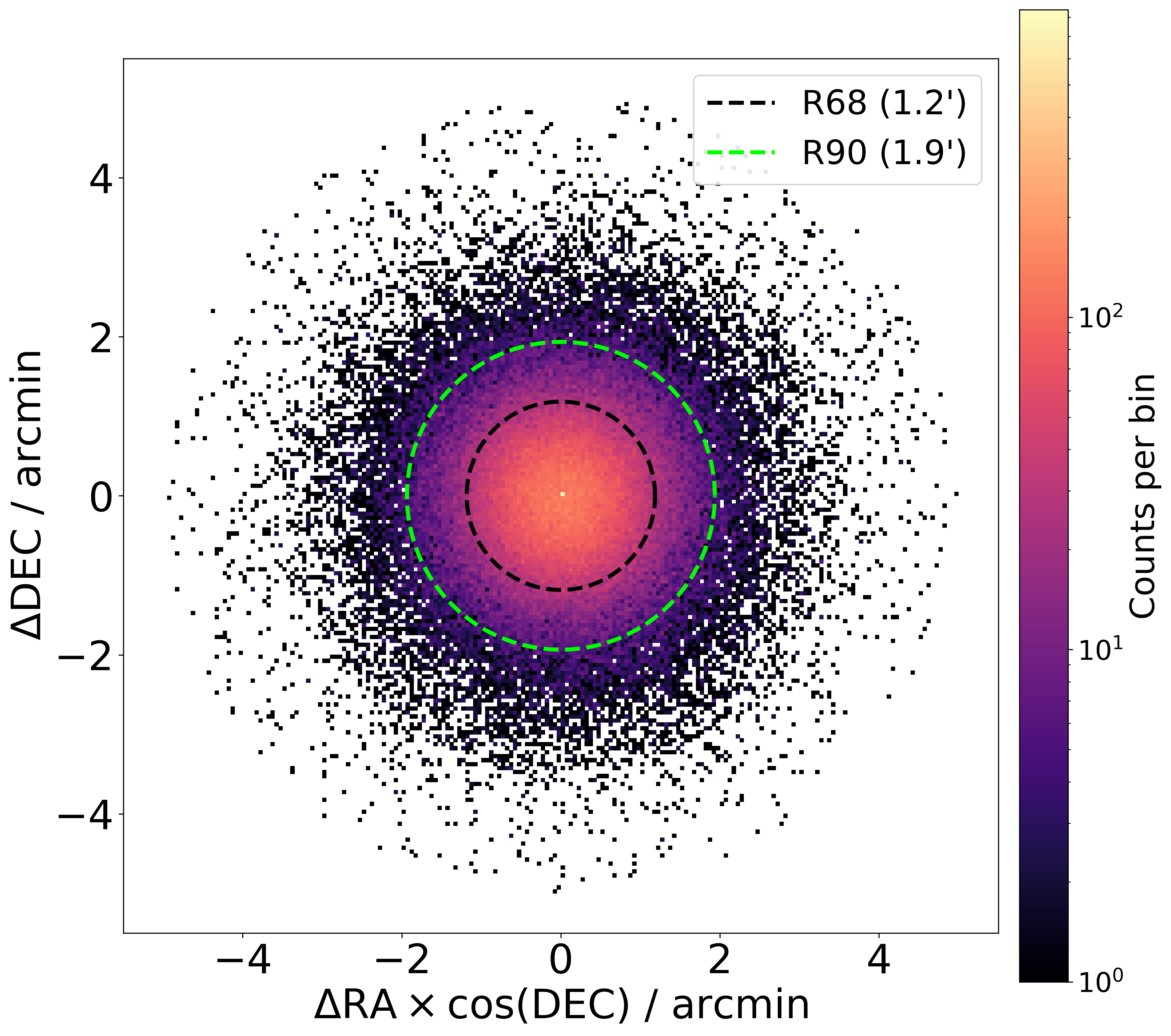}
    \caption{Offsets between the WXT positions and the NED coordinates for 164,525 valid detections from 4,370 sources, displayed as a two-dimensional histogram. The two circles denoted by dashed black and green lines correspond to the 68th and 90th percentiles of $1.2'$ and $1.9'$, respectively.}
    \label{fig:focalspot_offset}
\end{figure}

\begin{figure}[!htbp]
    \centering
    \includegraphics[width=\columnwidth]{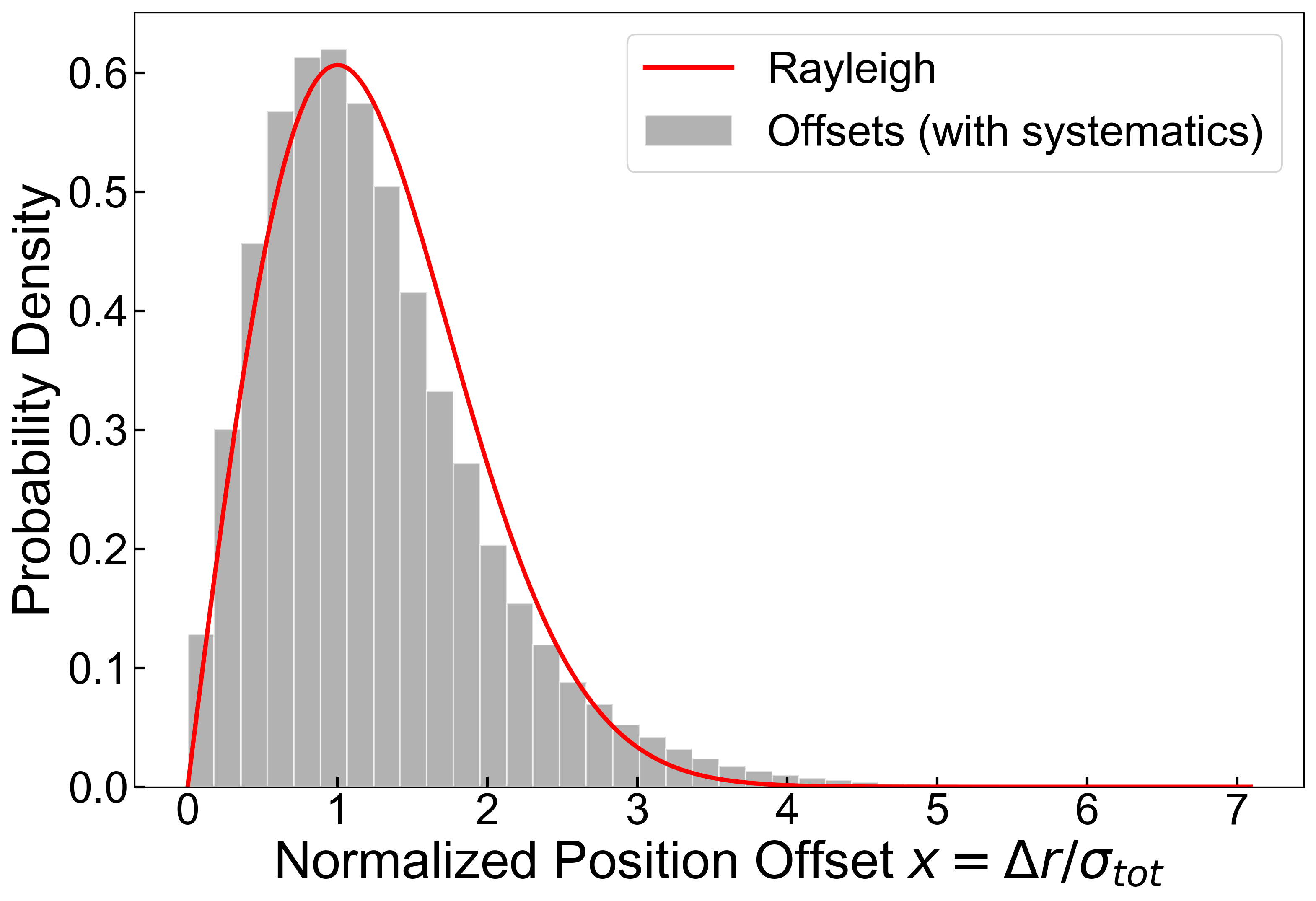}
    \caption{Histogram (in grey) of the normalized positional offset $x$ after incorporating the best-fit systematic uncertainty of $0.61'$ (which corresponds to $1.3'$ at the 90\% C.L.; see text for details), plotted alongside the theoretical Rayleigh distribution (red curve). A total of 164,525 detections from 4,370 sources are utilized for this calculation.}
    \label{fig:wxt_sys_error}
\end{figure}

\begin{figure}[!htbp]
    \centering
    \includegraphics[width=\columnwidth]{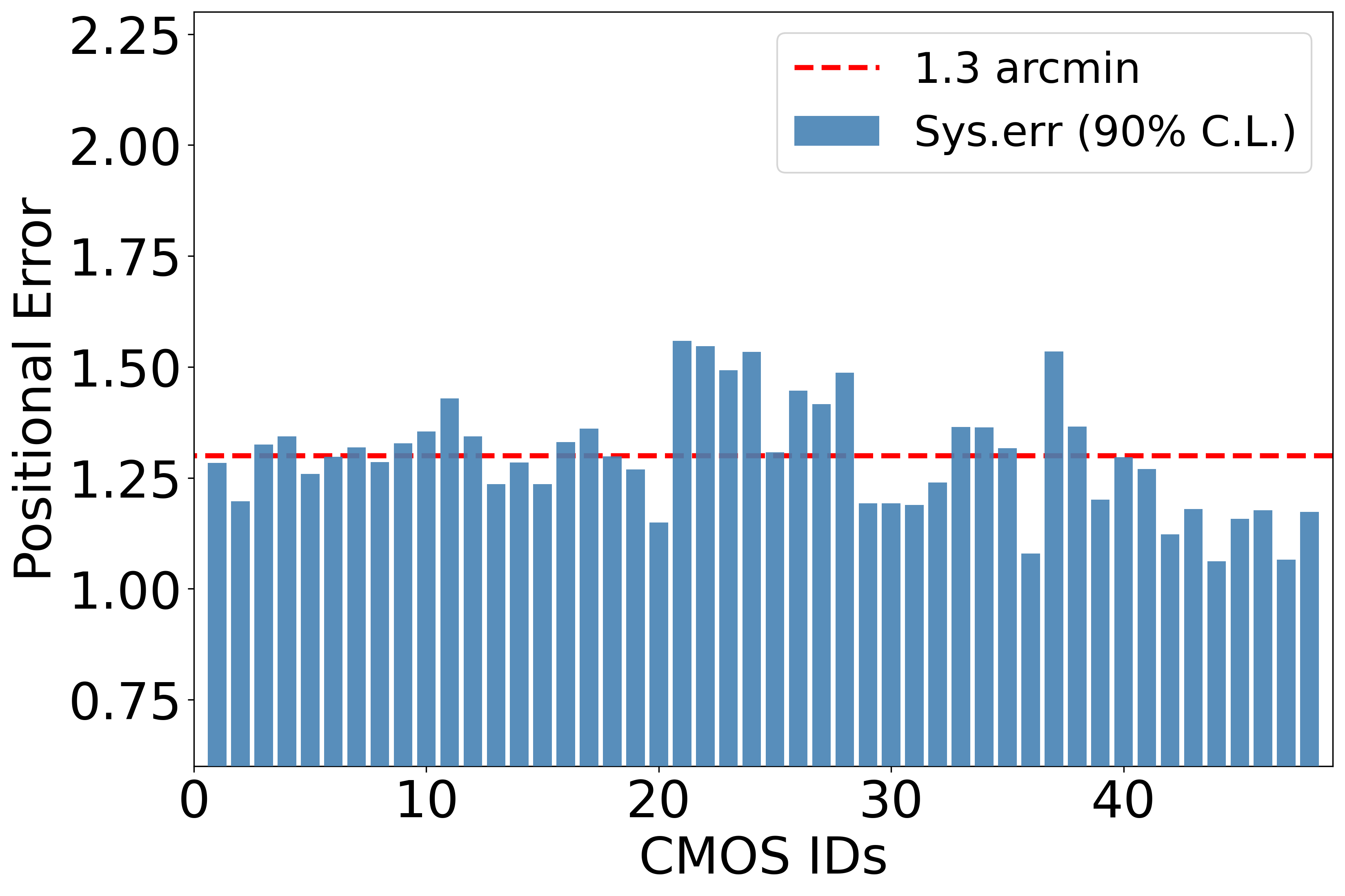}
    \caption{Systematic source positioning error (at the 90\% C.L.) for each individual CMOS detector. The red dashed line indicates the representative overall systematic uncertainty of $1.3'$ for the entire WXT instrument.}
    \label{fig:pos_err_cmos_distribution}
\end{figure}

The absolute source positional accuracy of the \textit{EP}-WXT was ultimately evaluated by comparing the detected source positions with their high-precision reference coordinates from the NASA/IPAC Extragalactic Database (NED)\footnote{\url{https://ned.ipac.caltech.edu/}}, utilizing the continuously collected post-calibration scientific data. For this purpose, we initiated our analysis using the WXT all-sky source catalog constructed from observational data acquired between August 1, 2024, and November 10, 2025 (UT), spanning over 15 months of in-orbit operations. 
The initial catalog contains 218,556 valid detections from 5,005 individual sources. These detections were subsequently screened in two steps to minimize potential biases that could skew the statistical assessment. 
First, 28,190 detections from 409 extended sources (e.g., supernova remnants, galaxy clusters, and star clusters) were rigorously removed. 
Subsequently, we discarded detections located near the detector boundaries (i.e., within 200 pixels from the edge), which amounts to 25,841 detections from 409 sources, as the PSFs in these regions are prone to distortion due to vignetting effects.
The final screened catalog comprises 164,525 detections from 4,370 point-like sources. 
Figure~\ref{fig:focalspot_offset} presents the two-dimensional histogram of the positional deviations for these screened detections. 
We find that the vast majority of the positional deviations ($\Delta r$) fall within $3'$, exhibiting a 68th percentile of $1.2'$ and a 90th percentile of $1.9'$.

With this massive dataset, we further estimated the systematic source positioning uncertainty, broadly following the standard approach detailed in \citet{Watson2009_possys}. 
Specifically, the normalized angular offset between the WXT source and its reference counterpart, defined as $x = \Delta r / \sigma_{\mathrm{tot}}$, is statistically expected to follow a Rayleigh distribution, $N(x)dx\propto x e^{-x^2/2}dx$ \citep[see e.g.,][for methodological details]{2012Feigelson}. Here, $\sigma_{\mathrm{tot}}$ represents the total positional error, calculated as the quadrature sum of the 1-dimensional statistical uncertainty ($\sigma_{\mathrm{1D}}$) and the intrinsic systematic error ($\sigma_{\mathrm{sys}}$), such that $\sigma_{\mathrm{tot}} = \sqrt{\sigma_{\mathrm{1D}}^2 + \sigma_{\mathrm{sys}}^2}$. By performing a maximum likelihood estimation (MLE) to match the observed offset distribution with the theoretical Rayleigh profile, we determined the best-fit $\sigma_{\mathrm{sys}}$ to be $\sim 0.61'$. Consequently, the total positional error radius at a 90\% confidence level can be strictly constrained as $R_{90} = 2.146\sigma_{\mathrm{sys}} \approx 1.3'$. As an illustration, the histogram of the normalized positional offset $x$, after incorporating the best-fit systematic uncertainty of $\sigma_{\mathrm{sys}}=0.61'$, is plotted in Fig.~\ref{fig:wxt_sys_error} alongside the theoretical Rayleigh distribution.

Because the 48 CMOS sensors are mosaicked to cover the entire FoV, we present the systematic error (at the 90\% C.L.) derived for each individual detector in Fig.~\ref{fig:pos_err_cmos_distribution}, to illustrate the spatial uniformity of the source localization accuracy. We find that, despite slight fluctuations driven primarily by minor variations in imaging quality (see, e.g., Fig.~\ref{fig:psf_fwhm_each_cmos}), the localization performance is highly consistent across different detectors. Every single CMOS sensor successfully meets the mission design requirement of $\le 2.0'$ at the 90\% C.L. Consequently, we confidently adopt $1.3'$ as the representative overall systematic localization uncertainty (90\% C.L.) for the entire \textit{EP}-WXT instrument.

It is worth noting that the source positioning accuracy of the \textit{EP}-WXT payload represents a significant improvement compared to that of its pathfinder mission, \textit{LEIA}, which exhibited a systematic uncertainty of $\sim 2.0'$ \citep{2025ChengLEIA}. 
This substantial performance enhancement is attributed fundamentally to the superior imaging quality of the flight model modules.

\section{Effective area}
\label{sec:effarea}

The calibration of the effective area for the WXT payload is performed through spectral analysis of the Crab nebula, which serves as a standard calibration source for the effective area of X-ray missions. This source exhibits a stable and nearly featureless X-ray continuum across the 1--100\,keV energy band, which can be well described by an absorbed power-law model with a photon index of $\Gamma\sim 2.1$, a hydrogen column density of $N_{\rm H}\sim 5\times10^{21}~{\rm cm^{-2}}$, and a normalization of $N\sim 10$ \citep[e.g.,][]{Toor_Seward_1974,2005SPIE.5898...22K,Weisskopf2010Crab,2015Mason_NuSTAR_calibration}.

The data from both the dedicated calibration observations and the regular scientific observations across 46 detectors were employed for the validation and calibration of the in-orbit effective area. We note that data from CMOS 34 and 41 were excluded from the subsequent analysis due to the previously discussed micrometeoroid impacts. 
Specifically, for the 39 detectors onboard ten FMs (FM 1--2, 4--8, and 10--12), the data from the scheduled calibration observations during the commissioning phase were utilized. 
For the remaining seven detectors (CMOS 9--12, 33, and 35--36) aboard FMs 3 and 9 (where no Crab observations were scheduled during the commissioning phase due to visibility limitations), scientific observational data obtained from October 2024 to March 2025 were utilized. These data were collected whenever the source entered the FoV of these detectors with a sufficiently long exposure time ($\ge 200$\,s).
For each observation, we analyzed the 0.4--8\,keV spectrum\footnote{The theoretical lower boundary of the energy band for the CMOS detectors is $\sim 350$\,eV, determined by the onboard event threshold configuration. The more conservative boundary of 0.4\,keV adopted in this work is reasonable, as the effective area decreases significantly toward the lowest energies.} using the \textsc{XSPEC} software \citep{1996ASPC..101...17A}. Following common practice, an absorbed power-law model (\texttt{tbabs*powerlaw} in \textsc{XSPEC}) is adopted. The \texttt{wilms} abundances \citep{2000wilms} and \texttt{vern} photoelectric cross-sections \citep{Vern1995xsect} are used throughout the analysis.

\begin{figure}[!htbp]
    \centering
    \includegraphics[width=\columnwidth]{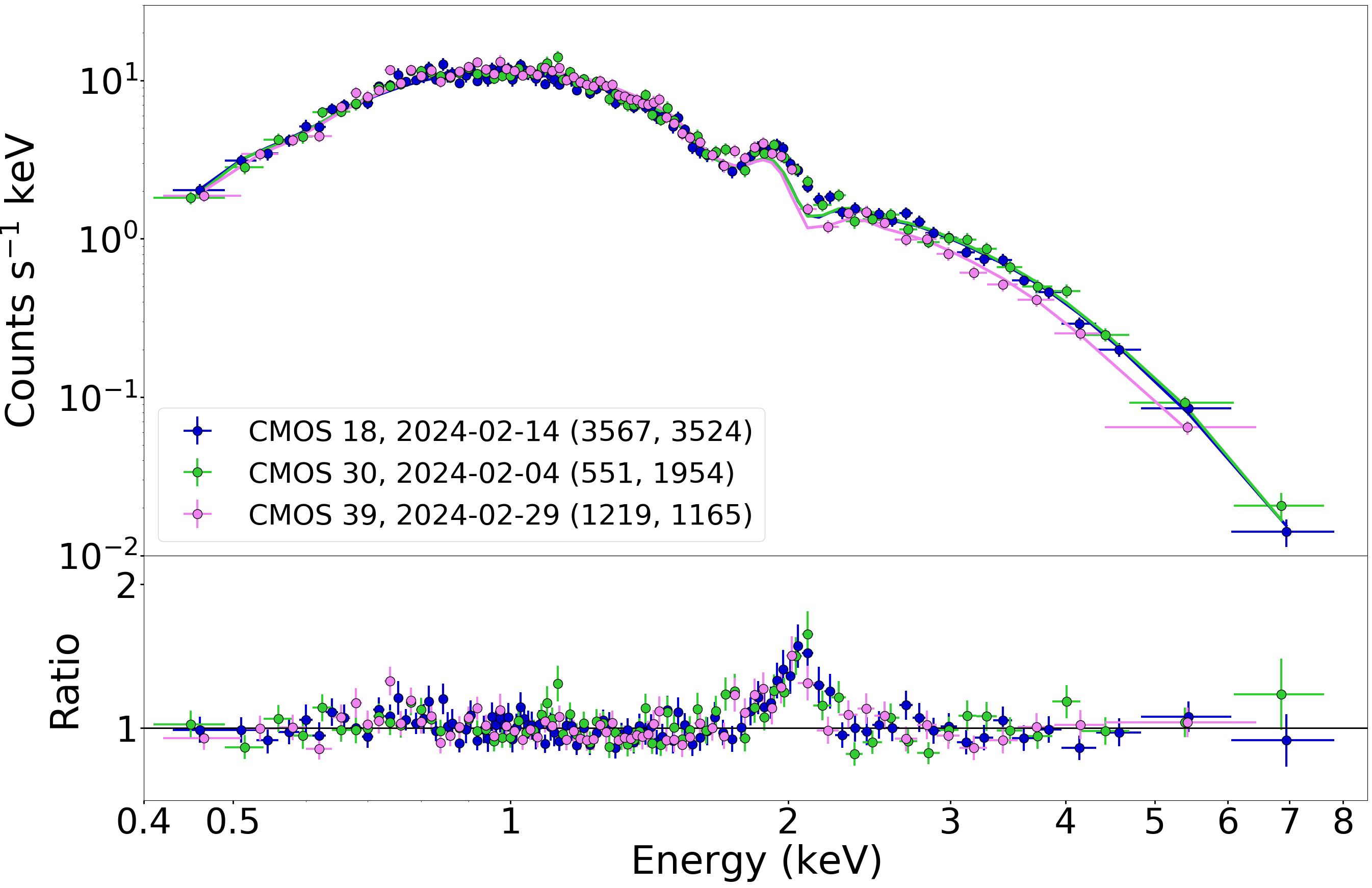}
    \caption{Spectral analysis of three Crab observations conducted during the commissioning phase. The source was located at the detector position of [3567, 3524] for CMOS 18 (blue symbols), [551, 1954] for CMOS 30 (green symbols), and [1219, 1165] for CMOS 39 (pink symbols). The data points were re-binned for illustrative purposes.}
    \label{fig:crab_spectralfit_normal_phase1}
\end{figure}

As an illustrative example, Fig.~\ref{fig:crab_spectralfit_normal_phase1} presents the spectral fitting results for three representative CMOS detectors (18, 30, and 39), each utilizing one observation taken during the commissioning phase (OBSIDs: 13600003659, 10200000288, and 10200001320). Specifically, in these observations, the Crab was positioned at different detector coordinates: [RAWX, RAWY] $\approx$ [3567, 3524], [551, 1954], and [1219, 1165], respectively. The \textit{upper} panel shows the spectra and the best-fit models, while the \textit{lower} panel displays the residuals.
All three spectra can be reasonably well reproduced by the conventional absorption-corrected power-law model, yielding best-fit statistics of $\chi^2/{\rm d.o.f.} = 246/231$, $265/207$, and $193/194$, respectively. 
The best-fitting column densities are $(4.3\pm 0.2)\times10^{21}$, $(4.4\pm 0.2)\times10^{21}$, and $(4.5\pm 0.2)\times10^{21}~{\rm cm^{-2}}$, respectively.
The best-fitting normalizations are $10.5\pm 0.5$, $10.7\pm 0.5$, and $10.0\pm 0.5$, respectively.
The best-fitting photon indices are $2.12\pm 0.05$, $2.10\pm 0.06$, and $2.07\pm 0.07$, respectively. 
The observed 0.5--4\,keV flux is approximately $2\times10^{-8}~{\rm erg~s^{-1}~cm^{-2}}$ for all three datasets.
The derived spectral parameters are thereby consistent among the different detectors within their mutual uncertainties.
Meanwhile, a noticeable residual around 2\,keV is present in all three spectra, exhibiting a data/model ratio of $\sim 1.5$--$2$, which is largely in agreement with the pattern observed in the pathfinder mission \textit{LEIA} \citep{2025ChengLEIA}. 
We suggest that this narrow-band spectral residual is likely a result of an imperfect modeling of the effective area across the Iridium absorption edge \citep[see also Sect.~5 of][for discussion]{2025ChengLEIA}. Detailed analysis indicates that the strength and width of this residual exhibit complex behavior, varying both among different detectors and across different positions on a single detector. 

\begin{figure*}[!htbp]
    \centering
    \includegraphics[width=\textwidth]{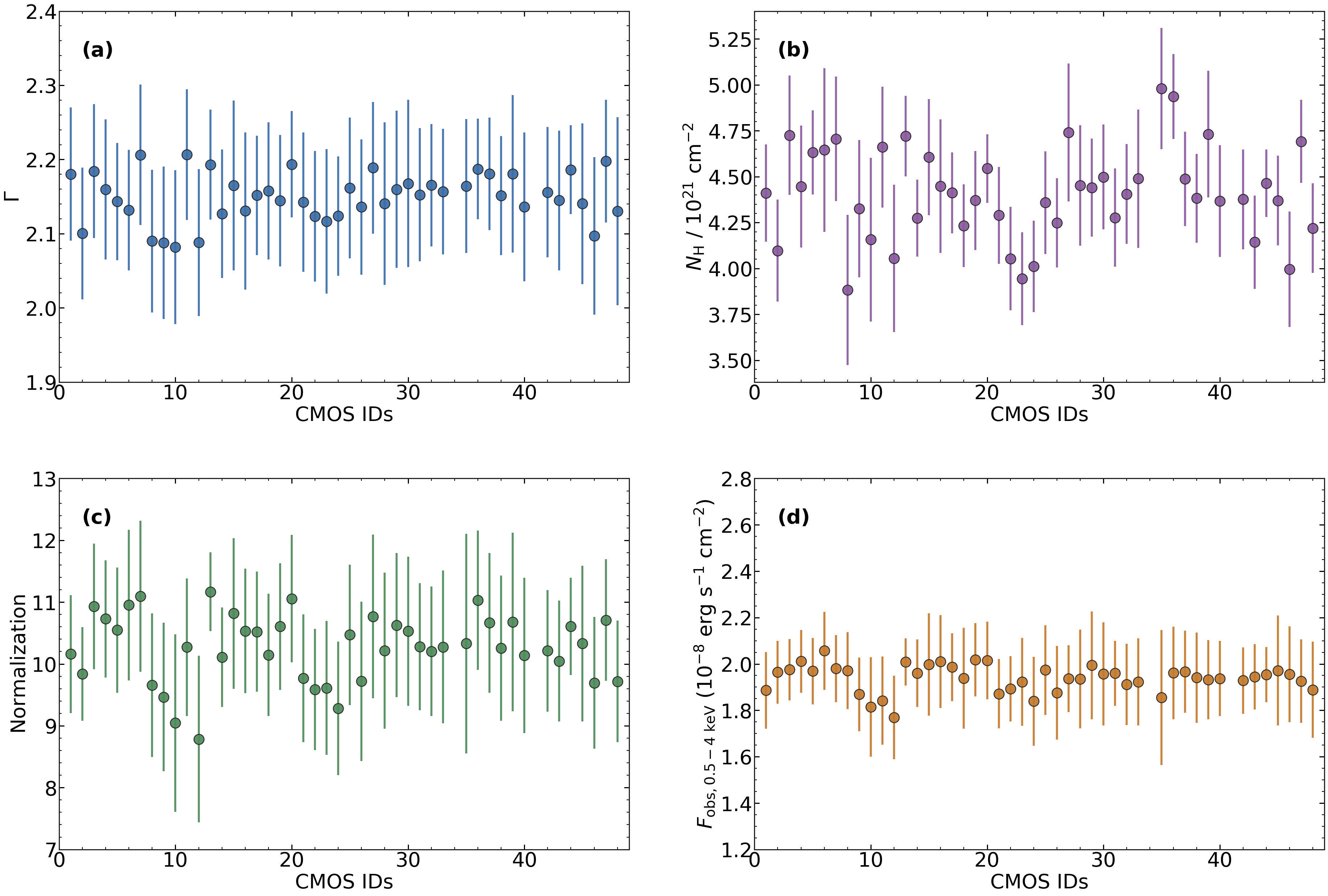}
    \caption{Crab spectral fitting parameters for 46 detectors (excluding CMOS 34 and CMOS 41) onboard the \textit{EP}-WXT instrument. Panels (a)--(d) show the photon index, the column density, the normalization of the power-law model, and the observed flux within the 0.5--4\,keV band, respectively. For each detector, the data point represents the typical parameter value calculated by averaging the best-fit results derived from different incident angles. The corresponding error bars denote the systematic scatter (at the 90\% C.L.) across the field of view, which is evaluated by isolating the intrinsic parameter dispersion from the statistical fitting noise. The observational data for the eight detectors onboard FM3 and FM9 were taken during the second calibration epoch. For the other 38 detectors, the observational data were taken during the commissioning phase. See the main text for details.}
    \label{fig:crab_spec_params_sum_46cmos}
\end{figure*}

We further investigated the systematic uncertainty of the modeled effective area and evaluated its response uniformity across different modules. Specifically, for each detector, we calculated the typical values of the three spectral parameters ($N_{\rm H}$, $\Gamma$, and the power-law normalization) alongside the $0.5$--$4$\,keV observed flux, which were determined by averaging the best-fit results obtained at different incident angles. To quantify the intrinsic systematic scatter across the field of view, we explicitly disentangled the underlying physical parameter dispersion from the statistical fitting noise. This was achieved by first calculating the total observed scatter ($\sigma_{\mathrm{obs}}$) as the sample standard deviation of the best-fit values. We then estimated the typical statistical uncertainty ($\sigma_{\mathrm{stat}}$) using the root-mean-square (RMS) of the individual $1\sigma$ fitting errors. The intrinsic systematic error ($\sigma_{\mathrm{sys}}$) was subsequently isolated via quadrature subtraction, namely $\sigma_{\mathrm{sys}} = \sqrt{\sigma_{\mathrm{obs}}^2 - \sigma_{\mathrm{stat}}^2}$. This derived systematic scatter, scaled to a 90\% confidence level, is presented as the error bars for each parameter in Fig.~\ref{fig:crab_spec_params_sum_46cmos}.

The main findings can be summarized as follows: 
(1) For an individual detector, the intrinsic systematic scatter for both the spectral shape parameters and the $0.5$--$4$\,keV observed flux is constrained to $\lesssim 10\%$. This physical dispersion provides a robust estimation for the systematic uncertainty of the modeled effective area.
(2) The fitting parameters and the measured fluxes are broadly consistent across the majority of the CMOS detectors, indicating a highly uniform absolute effective area among different modules. We note a marginal elevation in the column density for two specific detectors (CMOS 35 and 36), which is likely attributable to a mild deterioration of the effective area at the lower-energy end following approximately one year of in-orbit operations (see detailed discussion below).
(3) Taking these systematic uncertainties into account, the Crab spectral parameters derived by the \textit{EP}-WXT are in excellent agreement with the canonical values established by previous X-ray missions \citep[e.g.,][]{1997ApJ...491..808P,2000A&A...361..695M,2005SPIE.5898...22K,2022JATIS...8c4003M}.
Based on these results, we conclude that the in-orbit effective area of the WXT instrument agrees well with both model predictions and pre-flight ground calibration results, exhibiting an overall systematic uncertainty of $\lesssim 10\%$ ($90\%$ C.L.).

\begin{figure*}[!htbp]
    \centering
    \includegraphics[width=\columnwidth]{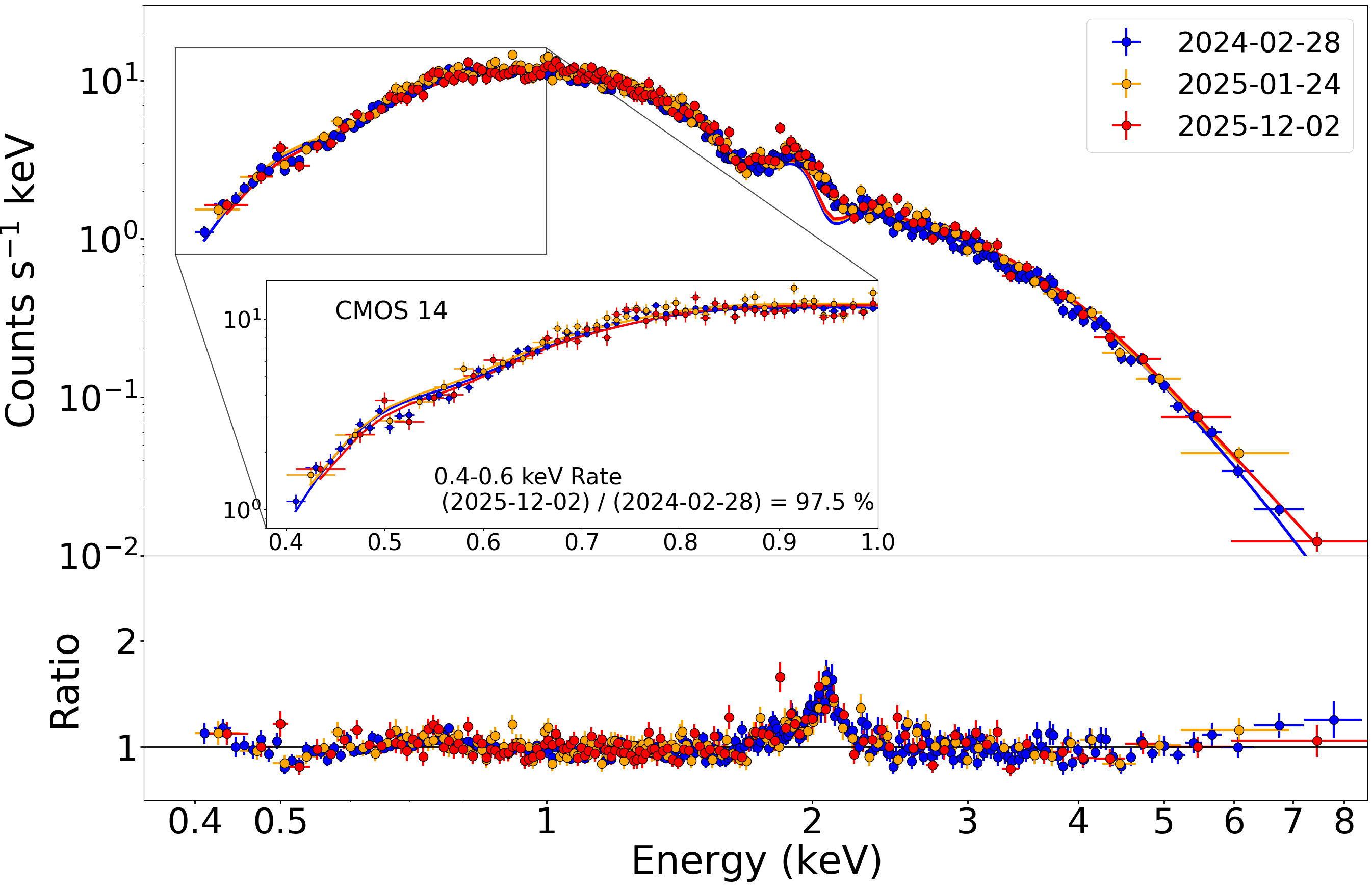}
    \includegraphics[width=\columnwidth]{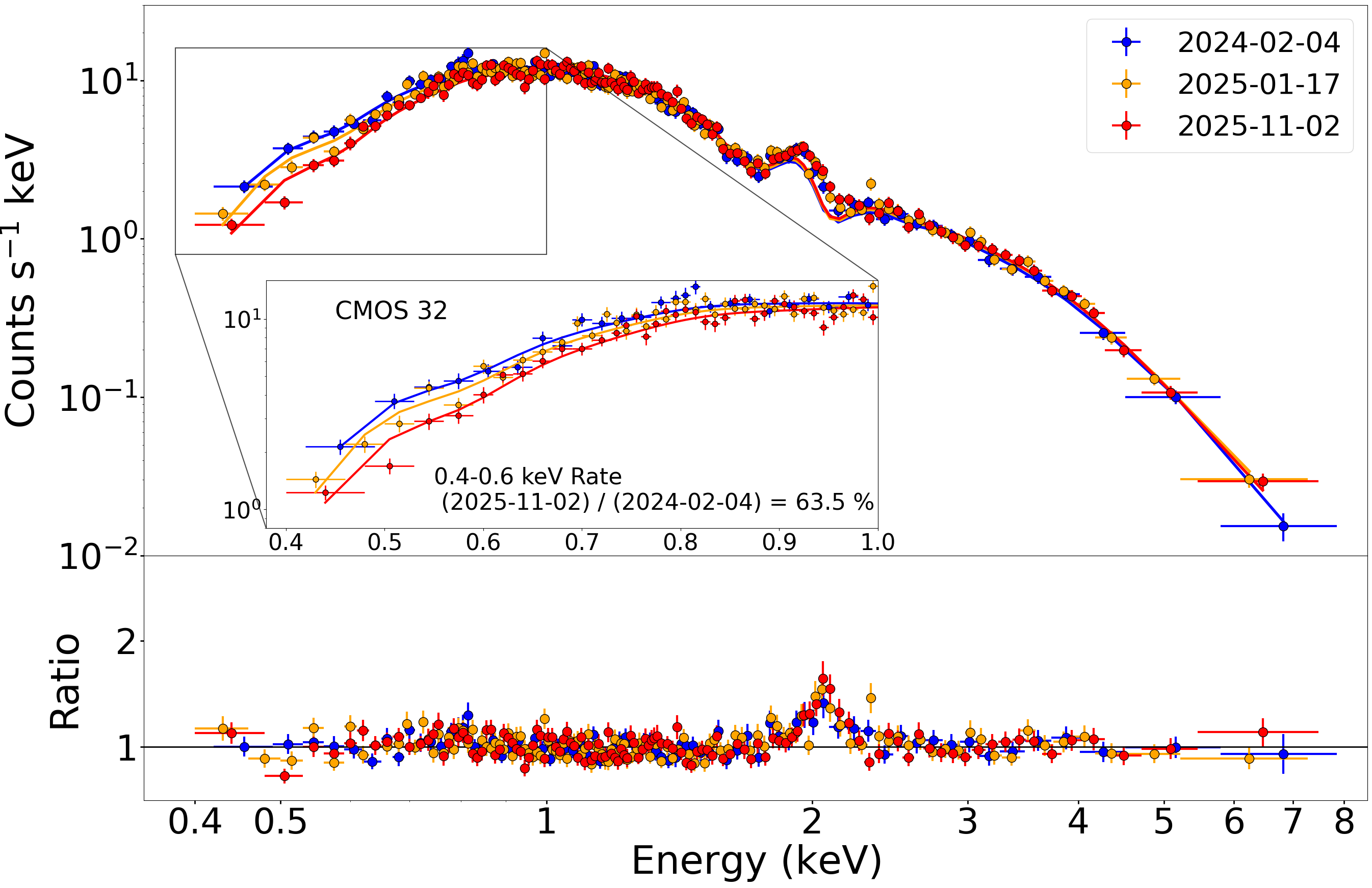}
    \caption{Crab spectra obtained on CMOS 14 ({\it left} panel) and CMOS 32 ({\it right} panel) during different calibration epochs: February 2024, January 2025, and November 2025. The source was located at the center of the detector in these observations. The lower-energy ends (0.4--1\,keV) of the spectra are zoomed in for clearer illustration.}
    \label{fig:eff_deter_cmos31_cmos32}
\end{figure*}

\begin{figure}[!htbp]
    \centering
    \includegraphics[width=\columnwidth]{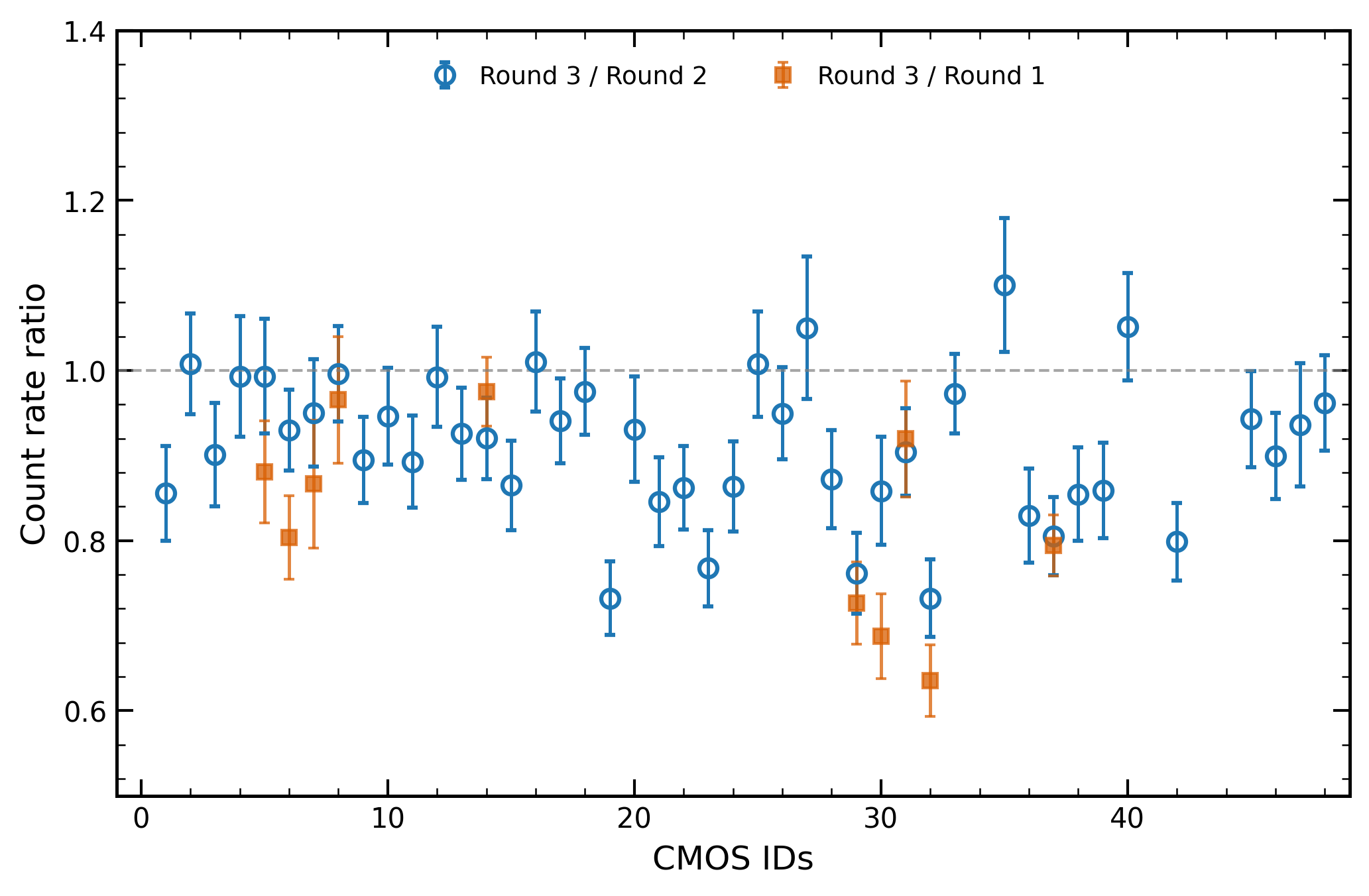}
    \caption{Ratio of the 0.4--0.6\,keV count rate across different calibration epochs. The red dots denote the ratio between the third epoch (data obtained from November 2025 to February 2026) and the second epoch (data obtained from January to February 2025), and the gray squares denote the ratio between the third epoch and the first epoch (data obtained from February to March 2024).}
    \label{fig:loweff_cmoslist}
\end{figure}

\begin{table*}[!htbp]
\caption{Summary of the spectral fitting results of the Crab nebula during the three calibration stages for CMOS 14 and CMOS 32. The \texttt{tbabs*powerlaw} model is utilized for all datasets, assuming \texttt{wilms} abundances and \texttt{vern} cross-sections.} 
\label{table:eff_deter} 
\centering                                        
\begin{tabular}{c c c c c c c c} 
\toprule
CMOS ID & OBSID & Observation date & $\Gamma$ & Norm & $N_{\rm H}$ & $\chi^2/\mathrm{d.o.f.}$ & Cts \\   
 & & & & & ($10^{21}$~cm$^{-2}$) & & (0.4--0.6~keV) \\ 
\midrule
\multirow{3}{*}{14} 
 & 08500000012 & 2024-02-28 & $2.11\pm0.02$ & $10.0\pm0.2$ & $4.3\pm0.1$ & $514/368$ & $0.54\pm0.01$\\
 & 11900063873 & 2025-01-24 & $2.14\pm0.05$ & $10.3\pm0.4$ & $4.4\pm0.2$ & $224/203$ & $0.57\pm0.02$\\
 & 13600497537 & 2025-12-02 & $2.11\pm0.05$ & $10.2\pm0.4$ & $4.5\pm0.2$ & $213/203$ & $0.53\pm0.02$ \\ 
\midrule
\multirow{3}{*}{32} 
 & 10200000290 & 2024-02-04 & $2.17\pm0.06$ & $10.6\pm0.5$ & $4.4\pm0.2$ & $230/204$ & $0.57\pm0.03$\\
 & 11900051073 & 2025-01-17 & $2.14\pm0.05$ & $10.7\pm0.4$ & $4.7\pm0.2$ & $345/247$ & $0.50\pm0.02$\\
 & 13600459521 & 2025-11-02 & $2.20\pm0.05$ & $11.8\pm0.5$ & $5.3\pm0.2$ & $302/249$ & $0.36\pm0.02$\\
\bottomrule
\end{tabular}
\end{table*}

The long-term degradation of the effective area at low energies, which is typically interpreted as resulting from the progressive accumulation of contaminants on the surfaces of the optical system and/or detector modules, has been reported in previous X-ray missions such as \textit{Chandra} \citep[e.g.,][]{Marshall2004,Plucinsky2022,Grant2024} and \textit{XMM-Newton} \citep[e.g.,][]{Kirsch_2005}.
For the \textit{EP}'s pathfinder mission \textit{LEIA}, a slight deterioration in the effective area on the order of $\sim 15\%$ was also observed after one year of operations \citep{2025ChengLEIA}.
To evaluate this potentially long-standing effect for the WXT payload, the spectral fitting results of the Crab obtained at different calibration stages were carefully compared. 
As an illustrative example, Fig.~\ref{fig:eff_deter_cmos31_cmos32} presents the spectral fitting results obtained from three calibration epochs, using on-axis observations of two representative detectors (CMOS 14 in the \textit{left} panel and CMOS 32 in the \textit{right} panel). 
In both panels, the \textit{upper} section shows the spectra and best-fit models, with a zoom-in view at the lower energy end (0.4--1.0\,keV) for clearer illustration, while the \textit{lower} section presents the residuals. The detailed fitting results are summarized in Table~\ref{table:eff_deter}. For CMOS 14, the spectral parameters are found to remain highly consistent across the three epochs. 
Neither a significant variation in the overall spectral shape nor a substantial decrease in the low-energy (0.4--0.6\,keV) count rate is observed ($C_{\rm 2025-12}/C_{\rm 2024-02}\approx 98\%$). In contrast, CMOS 32 exhibits a noticeable decreasing trend in the low-energy count rate (from $\sim 0.57~{\rm counts~s^{-1}}$ to $\sim 0.36~{\rm counts~s^{-1}}$, representing a moderate decrease of $\sim 37\%$), accompanied by increases in both $\Gamma$ and $N_{\rm H}$. This result explicitly indicates that the degree of low-energy effective area deterioration varies significantly among different detectors. 
Figure~\ref{fig:loweff_cmoslist} shows the ratio of 0.4--0.6\,keV count rates across separate calibration epochs for 44 CMOS detectors, using on-axis Crab observations\footnote{No on-axis Crab observations were conducted for CMOS 43 and CMOS 44 during the third calibration phase.}.
Specifically, the comparison between epoch-3 (spanning from October 2025 to February 2026) and epoch-1 (spanning from February to March 2024) can only be derived for 14 detectors (CMOS 5--8, 14, 29--32, and 37), as no on-axis observations were conducted for the other CMOS sensors during the commissioning phase. 
It is observed that, for certain detectors (e.g., CMOS 29, 30, and 32), the decrease in the low-energy count rate can be as severe as 30\%--40\% after $\sim 2$ years of operations, whereas other detectors (e.g., CMOS 5, 7, 8, 14, and 31) exhibit only a very slight effective area deterioration at the level of $\lesssim 10\%$.
On the other hand, the low-energy degradation measured within the second year of operation (as denoted by the red symbols) also exhibits a wide distribution among different detectors, ranging from a negligible level (e.g., CMOS 2, 4, 5, 8, 12, and 16) to a moderate level of $\sim 20\%$--$30\%$ (e.g., CMOS 19, 23, and 32).

The precise physical mechanism driving this diverse low-energy degradation across different detector modules remains under investigation. Given its complexity—potentially involving localized contamination, space weather effects, micrometeoroid impacts, or a combination thereof—dynamic correction factors have not yet been implemented in the current CALDB release. For the time being, it is advised to exercise caution when interpreting low-energy spectral features ($< 1$\,keV) originating from these severely affected modules. A comprehensive time-dependent modeling of the effective area is scheduled to be incorporated into future CALDB updates.

To summarize, the in-orbit effective area of the WXT payload is in good agreement with the theoretical expectations and ground calibration results.
A conservative estimation of its systematic uncertainty yields $\lesssim 10\%$ (at the 90\% C.L.).
The remaining residual structure around 2\,keV, likely associated with modeling uncertainties near the Iridium absorption edge, indicates that further refinements to the effective area curve are still required. 
During the initial two and a half years of on-orbit operations, a continuous decrease in the low-energy count rate of the Crab was observed for a few CMOS sensors.
Nevertheless, for the vast majority of the detectors, the decrements are much smaller. 
The moderate degradation of the low-energy effective area at the level of 30\%--40\% observed in specific modules necessitates dedicated calibration updates for these detectors and continued monitoring of all CMOS sensors in future calibration campaigns.

\section{Energy scale and spectral resolution}
\label{sec:cmos_energy}

\begin{figure}[!htbp]
    \centering
    \includegraphics[width=\columnwidth]{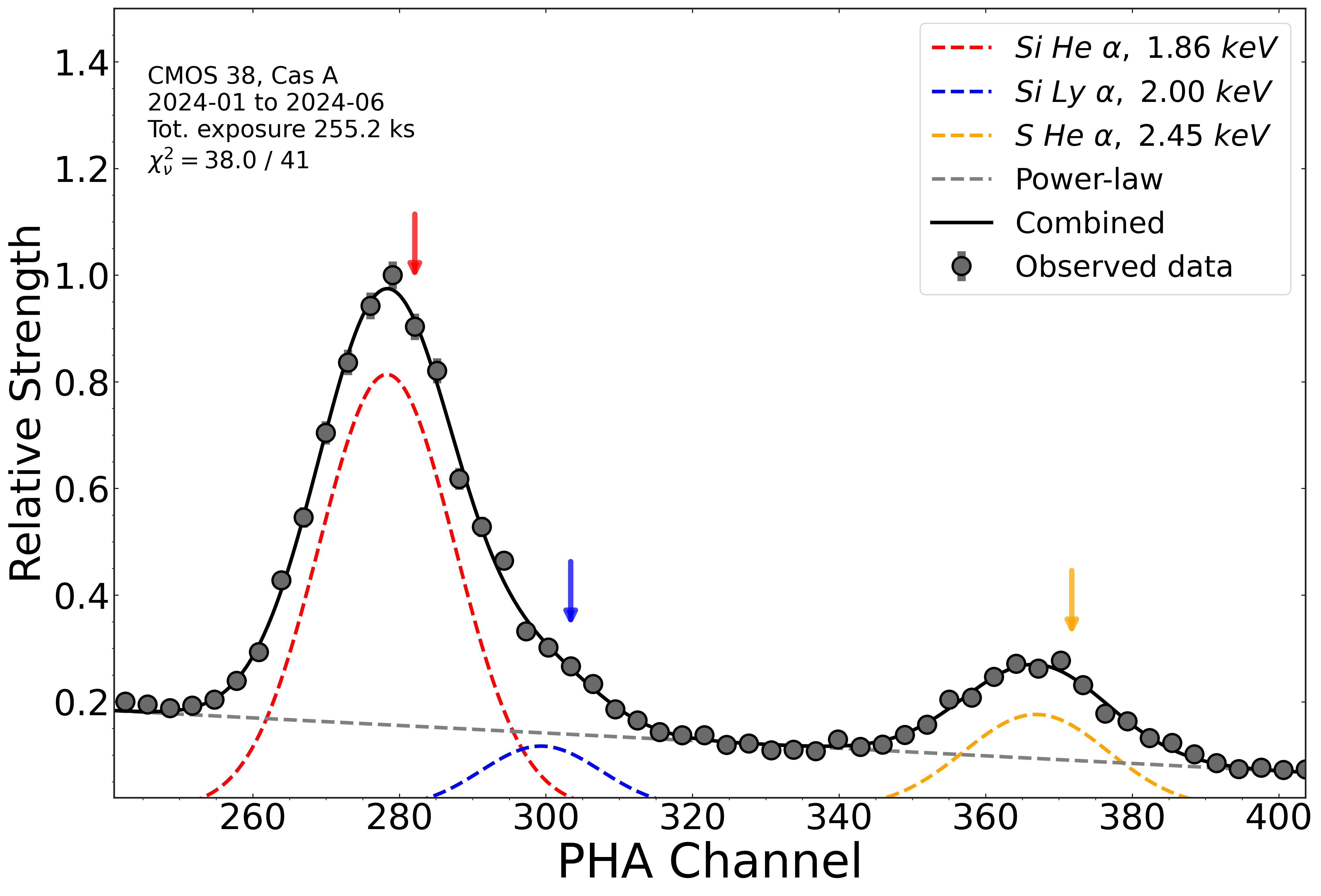}
    \includegraphics[width=\columnwidth]{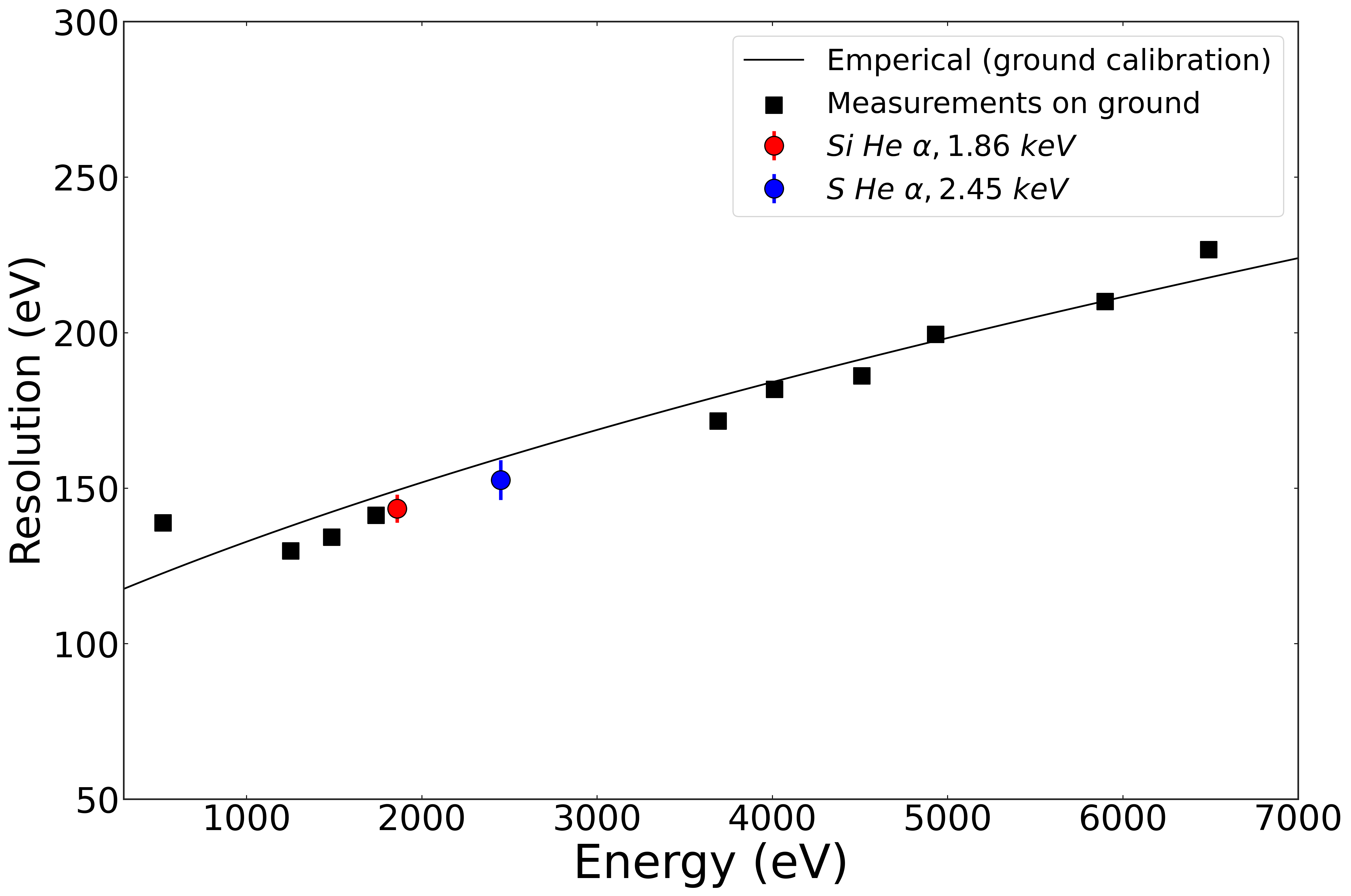}
    \caption{(\textit{Upper} panel) Spectral analysis of Cas A using the data retrieved from CMOS 38 during the commissioning phase, with a total exposure of $25.5$\,ks. The observed data are denoted by gray symbols. The underlying continuum, described by a power-law model, is signified by the gray dotted line. The three characteristic emission lines (Si He$\alpha$, S He$\alpha$, and Si Ly$\alpha$) are denoted by the red, blue, and orange dashed lines, respectively. The combination of these components is denoted by the black solid curve. The theoretically expected positions of the three emission lines, calculated from the ground-measured E-C relation, are marked by arrows in corresponding colors. (\textit{Lower} panel) Relation between the photon energy and the spectral resolution, derived from the Cas A spectral fitting for CMOS 38. The in-orbit spectral resolutions at the energies of Si He$\alpha$ and S He$\alpha$ are denoted by red and blue dots, respectively. 
    The resolutions measured from ground calibration experiments at ten characteristic X-ray emission lines 
    (including O K$\alpha$, Mg K$\alpha$, Al K$\alpha$, Si K$\alpha$, Ca K$\alpha$, Ca K$\beta$, Ti K$\alpha$, Ti K$\beta$, Mn K$\alpha$, and Mn K$\beta$)
    are denoted by black squares. The empirical relation obtained by fitting the standard resolution equation $\mathrm{FWHM}_{E}=2.35\omega(\sigma^2+f_{\rm Fano}E/\omega)^{1/2}$ \citep{Holland2013} to the ground measurements is denoted by the black solid curve.}
    \label{fig:casa_spectralfit_round1_cmos38}
\end{figure}

\begin{figure}[!htbp]
    \centering
    \includegraphics[width=0.49\textwidth]{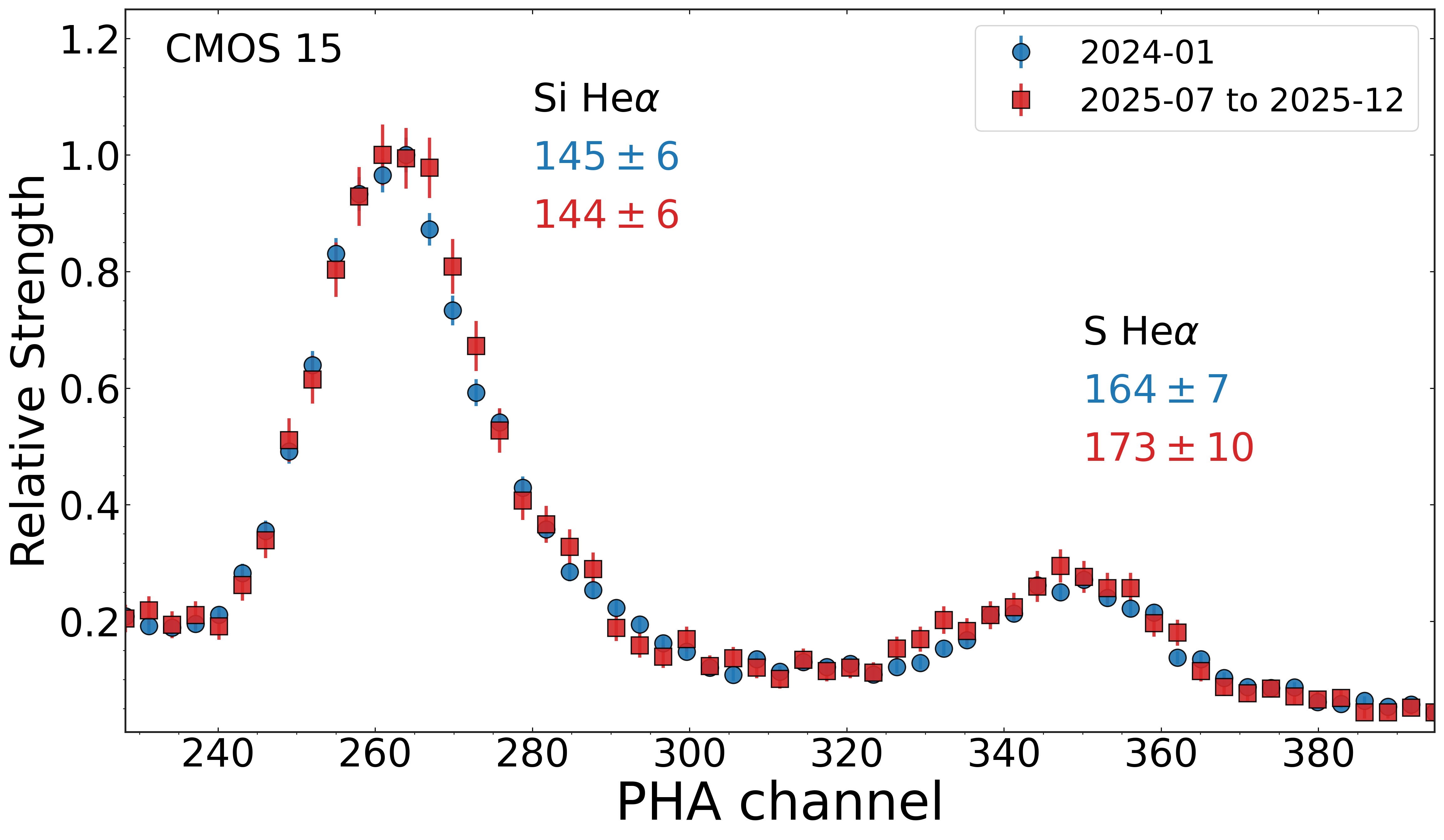}
    \caption{Comparison between the stacked Cas A spectra from different operational stages, utilizing observations from CMOS 15. The data represented by red dots were acquired during the commissioning phase, and those represented by blue dots were obtained during the third calibration epoch (approximately two years later).}
    \label{fig:casa_spec_compare_cmos38}
\end{figure}

For each of the 48 CMOS detectors aboard the \textit{EP} satellite, the energy response properties were rigorously measured at NAOC using a variety of targets (e.g., SiO$_2$, Al, Mg, Ca, Ti, and Mn) at different operating temperatures (Ling et al., in prep.). Furthermore, for the twelve detectors mounted on FMs 1, 5, and 11, additional calibrations were conducted at the 100-meter X-ray facility of IHEP using a multi-target electron impact X-ray source \citep[see Sect.~3.3 of][]{2025ChengWXTcalib}.
To establish the E-C relation and characterize the spectral resolution across the 0.4--8\,keV energy band, several primary X-ray emission lines were utilized in these ground calibration experiments, including K-shell lines (O K$\alpha$, Mg K$\alpha$, Al K$\alpha$, Si K$\alpha$, Ca K$\alpha$, Ca K$\beta$, Ti K$\alpha$, Ti K$\beta$, Mn K$\alpha$, and Mn K$\beta$) and L-shell lines (e.g., Cu L, Ag L$\alpha$). A highly linear E-C relation was observed, with only minor variations in the gain coefficient ($\sim 6.4$--$7$\,eV/DN) among the different detectors. The spectral resolution was typically measured to be approximately 120\,eV at 0.53\,keV, 130\,eV at 1.25\,keV, 140\,eV at 1.74\,keV, 160\,eV at 2.98\,keV, and 180\,eV at 4.51\,keV. Both the gain coefficient and the energy resolution exhibit weak spatial variations across individual detector planes ($<1\%$ and $<5\%$, respectively). These properties meet the design requirements well, providing a solid foundation for the reliable characterization of source emission properties.

During in-orbit operations, the Cas A supernova remnant was employed as a standard celestial calibrator to monitor potential post-launch variations in the energy response. 
To improve the signal-to-noise ratio, all available spectra for a given CMOS detector were stacked within each calibration phase, and adjacent PI channels were binned during the data analysis. 
During the commissioning phase, Cas A was primarily observed by a few detectors aboard FMs 4 and 10, given its short visibility window (before March 2024) and the tight schedule of the calibration campaigns for both the WXT and FXT instruments.
As an illustrative example, the stacked 1.6--2.7\,keV spectrum of Cas A obtained from CMOS 38 is presented in the \textit{upper} panel of Fig.~\ref{fig:casa_spectralfit_round1_cmos38}. Similar to the results obtained with the pathfinder \textit{LEIA} \citep[see Fig.~16 of][]{2025ChengLEIA}, three prominent emission lines---Si He$\alpha$ ($\sim 1.86$\,keV), Si Ly$\alpha$ ($\sim 2.0$\,keV), and S He$\alpha$ ($\sim 2.45$\,keV)---are clearly detected above the continuum. A phenomenological model comprising three Gaussian components superimposed on a power-law continuum was employed to fit the spectrum, yielding a statistically acceptable result ($\chi^2/{\rm d.o.f.} = 38/41$). The central PHA channels of the three emission lines agree remarkably well with the theoretical predictions based on the ground-calibrated E-C relation (denoted by colored arrows), showing offsets of only a few PHA channels. This explicitly indicates that the in-orbit E-C relation remains highly stable post-launch. The widths of the Gaussian components provide an accurate measure of the in-orbit spectral resolution near 2\,keV, yielding FWHM values of $143\pm 5$\,eV for Si He$\alpha$ and $153\pm 6$\,eV for S He$\alpha$. A direct comparison with the ground measurements (\textit{lower} panel of Fig.~\ref{fig:casa_spectralfit_round1_cmos38}) robustly demonstrates that the in-orbit spectral resolution is fully consistent with prelaunch expectations. The remaining CMOS sensors behave in a similar manner.

The large-scale deployment of CMOS sensors as focal-plane detectors is one of the most innovative technological explorations of the \textit{EP} mission.
During the ground experimental stage, their long-term stability was comprehensively investigated \citep[e.g.,][]{ChenMX2024, LiuMJ2025}.
Over the operational lifetime, we continue monitoring this stability by comparing Cas A spectra across different operational stages. As a representative example, Fig.~\ref{fig:casa_spec_compare_cmos38} shows the two Cas A spectra for CMOS 15, with one obtained during the commissioning phase (blue symbols) and the other from the third calibration stage (red symbols). 
The central channels and dispersions of the emission lines are found to be highly consistent across the two epochs, indicating a stable energy response even after $\sim 2$ years of in-orbit operations. 
We note that the results for the other CMOS sensors are similar. 
In fact, this energy response stability corroborates the findings of the pathfinder mission \textit{LEIA} \citep{2025ChengLEIA}, further validating that these customized scientific CMOS sensors exhibit superior radiation tolerance compared to traditional space-borne CCD detectors.

\section{Summary and conclusions}
\label{sec:summary}

The Wide-field X-ray Telescope (WXT) onboard the \textit{Einstein Probe} (\textit{EP}) mission represents a significant advancement in time-domain X-ray astronomy, being the first wide-field lobster-eye telescope with such an enormous instantaneous FoV ($> 3600$ square degrees) to complete a comprehensive in-orbit calibration campaign. This work details the results of these in-orbit calibrations conducted during the first two and a half years of operations, which are essential for validating the instrument's key performance parameters and ensuring the reliability of its scientific data products.

Overall, the in-orbit performance of the \textit{EP}-WXT is in good agreement with prelaunch ground calibrations and fully meets the scientific design requirements. Specifically, the PSF remains largely uniform across the expansive field of view, with the spatial resolution typically ranging from $3'$ to $6'$ (with a median of $\sim 4.3'$), exhibiting no discernible morphological degradation post-launch. The celestial-to-detector coordinate transformation matrices have been refined to achieve a highly reliable source positioning accuracy of $1.3'$ at the 90\% confidence level. This represents a significant improvement over the pathfinder mission \textit{LEIA} and comfortably satisfies the design goal of $\le 2.0'$, providing the high-precision localization necessary for prompt multi-wavelength follow-up observations of newly discovered transient sources.

The in-orbit effective area, characterized through extensive spectral analysis of the Crab nebula, is consistent with theoretical predictions and ground measurements, yielding an overall systematic uncertainty of $\lesssim 10\%$ (90\% C.L.). The vast majority of the CMOS detectors maintain a highly stable effective area over time; however, a few specific modules exhibit moderate degradation at the lower-energy end (30\%--40\%), highlighting the necessity for ongoing monitoring to track potential space-environment or contamination effects. Furthermore, the temporal evolution analysis of Cas A spectra robustly demonstrates that both the E-C relation and the spectral resolution agree well with ground calibrations and remain stable throughout the initial two and a half years of operations. 

In conclusion, the in-orbit calibration campaigns initiated during the first two and a half years of operations have thoroughly characterized the instrumental baseline and performance of the \textit{EP}-WXT payload. These results validate the instrument's innovative design and establish a solid foundation for achieving its primary scientific objectives. The calibration methodologies developed and refined in this work not only guarantee the reliable interpretation of WXT scientific data but also provide an invaluable reference for the design and operation of future wide-field X-ray observatories.

\begin{acknowledgements}
This work is based on the data obtained with the Einstein Probe (EP), also known as ‘Tianguan’. EP is led by the Chinese Academy of Sciences, in collaboration with the European Space Agency, the Max Planck Institute for Extraterrestrial Physics (Germany), and the Centre National d’Études Spatiales (France). This work is supported by National Key R\&D Program of China No.2025YFF0511100. This work is supported by the National Natural Science Foundation of China (Grant
Nos. 12333004, 12433005, 12473100, 12273073), and the Strategic Priority Research Program of the Chinese Academy of Sciences (Grant No.XDB0550200).
\end{acknowledgements}

\bibliography{ref}
\bibliographystyle{aa}

\end{document}